\documentclass[twocolumn]{aastex63}
\usepackage{graphicx}
\usepackage{natbib}
\usepackage{placeins}
\usepackage{amsmath}
\usepackage{tgcursor}
\usepackage{booktabs}
\usepackage{enumitem}
\usepackage{footnote}
\usepackage[version=4]{mhchem}
\setcitestyle{notesep={ }}

\newcommand{\tgrow}[0]{\tau_{\rm grow}}

\begin{document}

\title{Chemical Feedbacks of Pebble Growth: Impacts on CO depletion and C/O ratios}

\author[0000-0002-5954-6302]{Eric Van Clepper}
\affiliation{University of Chicago Department of the Geophysical Sciences, Chicago, IL 60637, USA}

\author[0000-0002-8716-0482]{Jennifer B. Bergner}
\affiliation{University of Chicago Department of the Geophysical Sciences, Chicago, IL 60637, USA}
\affiliation{NASA Sagan Fellow}

\author[0000-0003-4001-3589]{Arthur D. Bosman}
\affiliation{Department of Astronomy, University of Michigan, 323 West Hall, 1085 S. University Avenue, Ann Arbor, MI 48109, USA}

\author{Edwin Bergin}
\affiliation{Department of Astronomy, University of Michigan, 323 West Hall, 1085 S. University Avenue, Ann Arbor, MI 48109, USA}

\author{Fred J. Ciesla}
\affiliation{University of Chicago Department of the Geophysical Sciences, Chicago, IL 60637, USA}

\begin{abstract}
\noindent
Observations of protoplanetary disks have revealed them to be complex and dynamic, with vertical and radial transport of gas and dust occurring simultaneously with chemistry and planet formation. Previous models of protoplanetary disks focused primarily on chemical evolution of gas and dust in a static disk, or dynamical evolution of solids in a chemically passive disk. In this paper, we present a new 1D method for modelling pebble growth and chemistry simultaneously. Gas and small dust particles are allowed to diffuse vertically, connecting chemistry at all elevations of the disk. Pebbles are assumed to form from the dust present around the midplane, inheriting the composition of ices at this location.  We present the results of this model after 1 Myr of disk evolution around a 1$M_\odot$ star at various locations both inside and outside of the CO snowline. We find that for a turbulent disk ($\alpha = 10^{-3}$), CO is depleted from the surface layers of the disk by roughly 1-2 orders of magnitude, consistent with observations of protoplanetary disks. This is achieved by a combination of ice sequestration and decreasing UV opacity, both driven by pebble growth.  Further, we find the selective removal of ice species via pebble growth and sequestration can increase gas phase C/O ratios to values of approximately unity. However, our model is unable to produce C/O values of $\sim$1.5-2.0 inferred from protoplanetary disk observations, implying selective sequestration of ice is not sufficient to explain C/O ratios $>1$.
\end{abstract}


\section{Introduction}

Planets form in the cold, dense midplanes of accretion disks around nascent stars.  Recent observations of protoplanetary disks have provided a wealth of constraints on the physical and chemical processes associated with planet formation. Interferometric observatories, such as the Atacama Large Millimeter/submillimeter Array (ALMA), have probed the chemical compositions of the gaseous disks \citep{2021arXiv210906268O} and physical structures of the small dust \citep{2018ApJ...869L..41A} at high angular resolution. Such observations have revealed that the chemical inventories in disks are markedly different from those seen in molecular clouds, indicating significant processing in the solids and gas that eventually are incorporated into planets.  Notably, the CO abundance observed across protoplanetary disks is consistently found to be a factor of 10--100 lower than the canonical interstellar medium (ISM) value of $\sim$10$^{-4}$ \citep{2013ApJ...776L..38F, Schwarz2016, 2017A&A...599A.113M, 2019ApJ...883...98Z, Zhang2021}.  Additionally, the bright emission of hydrocarbons in disks points to significantly elevated gas-phase C/O ratios ($>$1) compared to the interstellar value \citep[$\sim$0.4;][]{2016ApJ...831..101B, 2018ApJ...865..155C, 2019AA...631A..69M, 2021ApJ...910....3B}.  
Understanding the origin of these chemical patterns is critical to understanding what sets compositions of incipient planetesimals and planets that are forming in these disks.

Protoplanetary disks are dynamic objects, within which a variety of chemical and physical processes are at work to transform the raw materials inherited from molecular clouds into the building blocks of planets.  The process of planet formation begins with the coagulation of dust grains, and their subsequent redistribution throughout the disk via settling and radial drift.  This leads to constantly changing physical conditions within the disk (e.g.~ density, temperature, UV fields), which in turn drives an evolving disk volatile chemistry \citep[e.g.][]{2020ApJ...898...97B}.  Thus, planet forming environments are characterized by co-evolving dynamics, physics, and chemistry.

Various models have attempted to explain the low observed CO abundances in disks, generally invoking either physical or chemical processes.  Physical models of CO removal consider sequestration of CO ice around the midplane: as small dust grains grow into larger pebbles, they settle to the midplane and trap the ice reservoirs in larger bodies \citep{2016A&A...592A..83K, 2020ApJ...899..134K}.  Such models can explain a factor of $\sim$10 decrease in the gas-phase CO abundance over $\sim$3 Myr timescales.  Chemical models of CO removal account for the processing of CO into larger, less volatile organics in the photon-dominated and warm molecular layers of the upper disk \citep{2015A&A...579A..82R, 2018IAUS..332...69E, 2018ApJ...856...85S}.  These models can also  achieve a factor of 10 depletion in CO for high cosmic ray ionization rates, but only in cold disk models.  In both cases, the depletion levels are on the low end of what is observed in disks, and on longer timescales ($\sim$3 Myr) than suggested by observations  \citep[$\sim$0.5--1 Myr][]{2020ApJ...898...97B, 2020ApJ...891L..17Z}.  As considering dynamics and chemistry independently cannot reproduce observations, more recent models have considered the combined effects, though so far only in simplified ways due to the computational demand of such efforts.  For instance, \citet{2020ApJ...899..134K} explored CO depletion in a dynamic disk with a simplified chemistry of CO hydrogenation on grains.

Chemical and dynamical effects may also be responsible for producing high gas phase C/O ratios within a protoplanetary disk. 
To achieve high C/O ratios in the gas via chemistry alone requires C to be incorporated into volatile, gaseous species while O is incorporated into solids.   Correspondingly,  selective sequestration of oxygen bearing ices (e.g. H$_2$O) via pebble growth and settling has been invoked as a means of producing a carbon-rich gas \citep{2016ApJ...831..101B, 2020ApJ...891L..17Z, 2021arXiv210906221B}. In this scheme, chemical destruction of CO is followed by formation and sequestration of H$_2$O, leading to an excess of carbon in the gas over the starting value.  Alternate sources of carbon exist, such as photoablation of small carbonaceous grains \citep[e.g.][]{Lee2010, Anderson2017, 2021ApJ...910....3B}; but these also rely on the presence of UV photons to power the chemistry.
Thus, pebble growth and chemical processing may both be necessary to reach high C/O ratios.

To date, few models have explored the detailed feedbacks that exist between pebble growth and chemistry in protoplanetary disks. \citet{2011ApJS..196...25S}, \citet{2013ApJ...779...11F}, and \citet{ 2014ApJ...790...97F},  have studied chemistry in a turbulent disk, but not including dust dynamics.  \citet{2019MNRAS.487.3998B} combined chemistry with pebble dynamics, but focusing only on the radial movement of large pebbles at the midplane. However, the growth of pebbles and removal of small dust will have an effect on chemistry throughout the entire column of the disk.  Because small dust is the primary source of UV opacity, its removal leads to greater penetration of UV radiation in the disk, initiating photochemical reactions deeper in the disk.  Thus, as grain growth proceeds, the removal of ices and the increasing UV penetration will lead to changing physical conditions and initiation of new chemical pathways  throughout a vertical column of the disk over time.

In this paper, we present a new model including a full chemical network along with pebble growth and ice sequestration, within a 1D vertical slice of a protoplanetary disk.  With this model, we explore the feedback of small dust removal and ice sequestration on gas phase chemistry as these processes occur concurrently.  Here, we focus on the effects of the coupled chemistry and dynamics on CO depletion and elevated C/O ratios. Previous studies have linked these features of protoplanetary disks to both chemical processing and dust evolution, and recent observations have shown that these outcomes are robust across protoplanetary disks \citep{2021arXiv210906221B,  2019ApJ...883...98Z}. In Section 2 we describe the disk structure and model details used for this paper, and compare the results of our new model with static and diffusive disk models in Section 3. Sections 4 and 5 examine the effects of pebble growth on CO depletion and C/O ratios respectively. Finally, a discussion of our results in Section 6 compares our findings to those from previous modeling efforts, and describes how these results should be used to interpret  disk observations.

\section{Methods}\label{methods}

\begin{figure}
    \centering
    \includegraphics[width=0.98\linewidth]{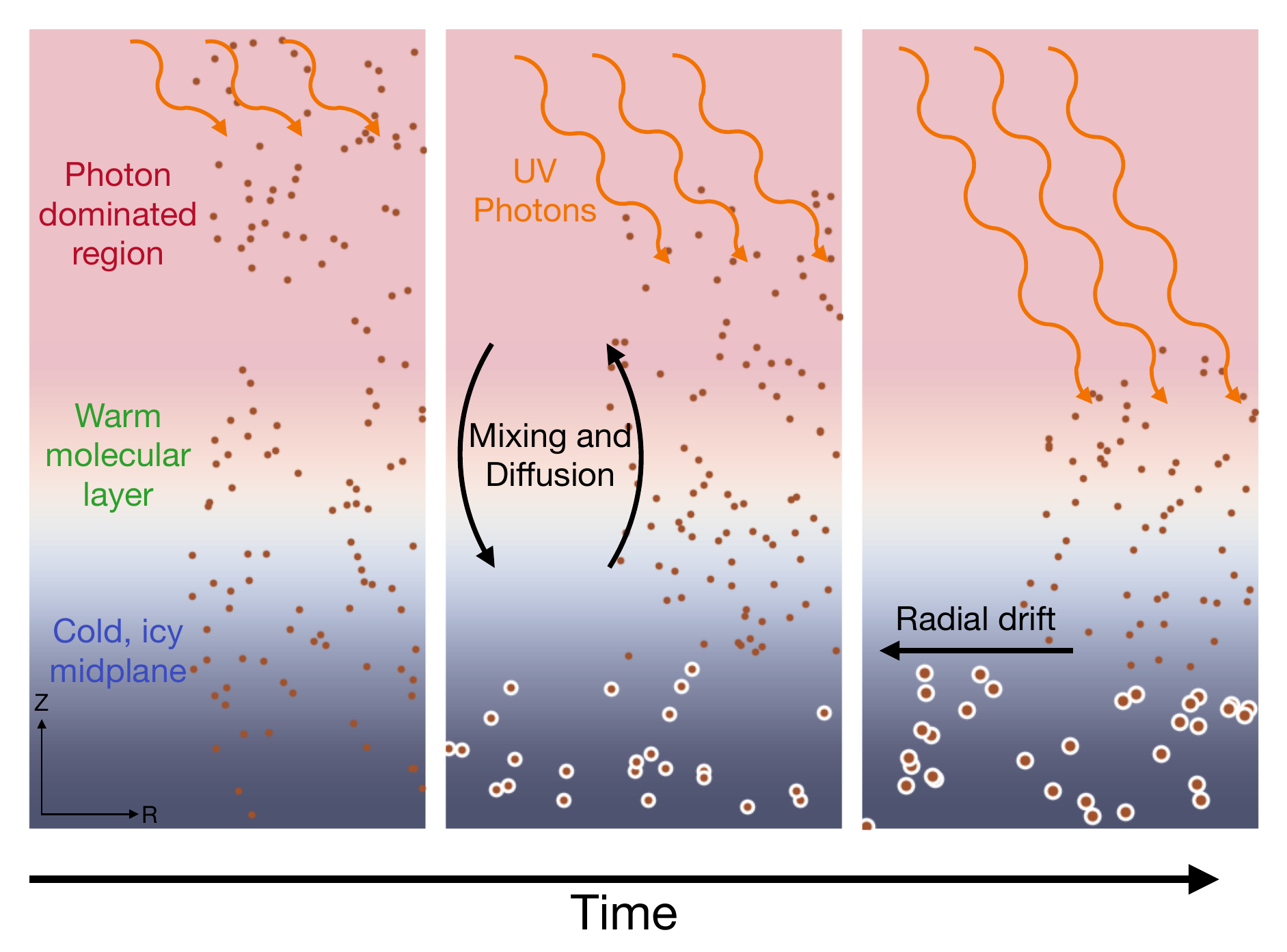}
    \caption{A cartoon diagram showing some of the feedbacks between pebble growth and gas phase chemistry. Early in the disk lifetime (left), small dust is well coupled with the gas at all elevations throughout a vertical slice of the protoplanetary disk. UV radiation penetrates into the uppermost layers of the disk in the PDL, but are attenuated by the small dust. At later stages (middle), small dust has begun to form into larger pebbles. These pebbles inherit the icy composition of the pebbles from which they form. As pebbles form, dust grains settle towards the midplane, allowing radiation to penetrate deeper into the disk. As the disk continues to age (right) pebbles drift out of the vertical column considered, taking their icy mantles with them. This acts to selectively remove ice phase species from the column, and combined with increased UV radiation can significantly alter the chemistry in the disk.}
    \label{fig:column_cartoon}
\end{figure}

Fig.~\ref{fig:column_cartoon} illustrates the physical processes investigated in our model.  We begin with a vertical column of the disk, in which fine dust is well-mixed with the gas at all heights (Fig~\ref{fig:column_cartoon}, left).  Chemistry occurs at all elevations as determined by the local conditions: the temperature, density, and local UV flux. Dust and gas is able to diffuse throughout the column, being exposed to the different conditions in the column, rather than staying fixed in place (Fig~\ref{fig:column_cartoon}, middle).   Pebbles are created around the disk midplane via coagulation of the dust (Fig~\ref{fig:column_cartoon}, right).  The creation of the pebbles removes dust from this region of the disk, which is then resupplied by diffusion from the upper layers.  As pebbles are created, the abundance of dust in the column decreases, allowing ambient radiation to penetrate deeper into the disk over time.  

In this section, we describe in detail the various components of this framework and how they are treated in our model.  We begin by describing the physical environment within which the processes are considered to operate.  We then describe how we treat the dynamical evolution of the gas and dust, and the assumptions made around the growth of pebbles.  We then outline the chemical model and network used in this study.  Finally, we describe how the various components are integrated to evaluate the chemical evolution that occurs within the described framework.

\subsection{Disk Structure}\label{disk_structure}

Because our model is 1D, we explore how these processes work across the disk by modeling vertical slices at various disk radii. We define a disk structure that is similar to \citet{2018ApJ...864...78K} for a disk around a 1 $M_\odot$ star. The radial surface density is given by

\begin{equation}
    \Sigma_g(r) = \Sigma_c\left(\frac{r}{r_c}\right)^{-p} \exp\left[-\left(\frac{r}{r_c}\right)^{2-p}\right],
\end{equation}
normalized such that

\begin{equation}
    \Sigma_c = (2-p)\frac{M_{\rm disk}}{2\pi r_c^2}.
\end{equation}
For our model, we choose $M_{\rm disk} = 0.05 M_\odot$, $p=1$, and $r_c = 100$ au. 
The midplane temperature follows the power law:

\begin{equation}
    T_{\rm mid}(r) = 130\left(\frac{r}{\rm{au}}\right)^{-1/2}\rm{K}.
\end{equation}

At any given distance from the star, $r$, the vertical density distribution is given by:

\begin{equation}
    \rho_g(r,z) = \frac{\Sigma_g(r)}{\sqrt{2\pi} H} \exp\left[-\frac{1}{2}\left(\frac{z}{H}\right)^2\right] ,
\end{equation}
where $z$ is the height above the disk midplane, $H = c_s/\Omega$ is the scale height of the gas and

\begin{equation}
    c_s = \sqrt{\frac{k_B T_{\rm mid}(r)}{\mu m_H}}
\end{equation}
is the sound speed for a given midplane temperature, $T_{\rm mid}$ with a mean molecular weight $\mu = 2.3$, $\Omega = \sqrt{GM_\odot/r^3}$ is the Keplarian orbital frequency and $k_B$ is the Boltzmann constant.

\begin{figure}
    \centering
    \includegraphics[width=\linewidth]{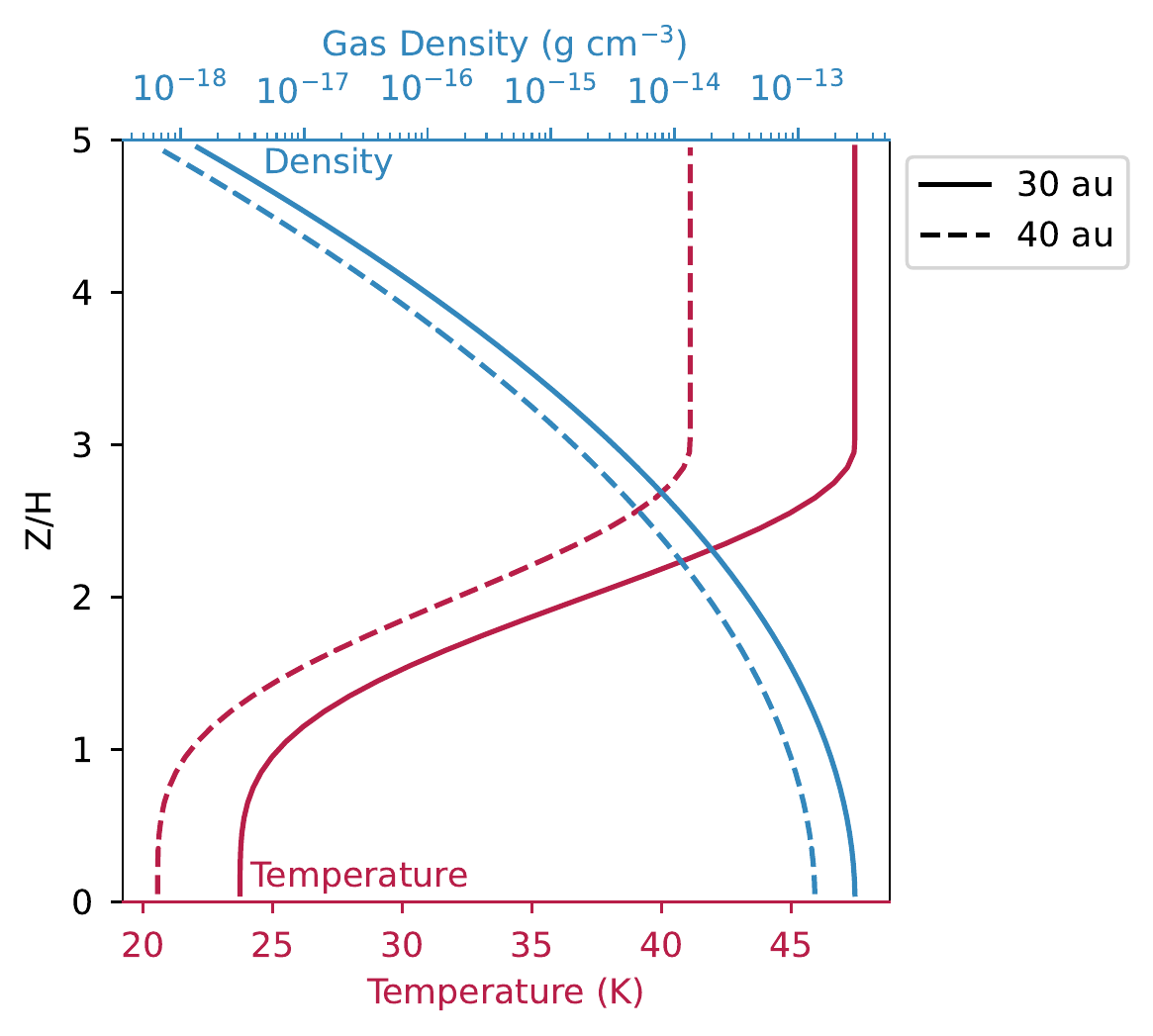}
    \caption{Temperature and density structure of the considered disk at 30 and 40 au. CO begins to freeze out at 21 K, so the midplane CO snowline lies between these two radii. }
    \label{fig:phys_structure}
\end{figure}

The surface layers (higher altitudes) of the disk are directly heated by radiation from the star, and thus reach higher temperatures than the midplane, which is only heated by reprocessed radiation.  To describe this structure, we follow \citet{2003A&A...399..773D} and set the temperature in the surface layers as twice that of the midplane.  That is:

\begin{equation}
    T(r,z) = T_{\rm atm}(r) = 2T_{\rm mid}(r).
\end{equation}
above a height $z=z_q H$.  Below this height, the temperature is given by:

\begin{equation}
    T(r,z) = T_{\rm mid}(r)\left(1 + \left[\sin\frac{\pi z}{2z_qH}\right]^{2\delta}\right),
\end{equation}

\noindent where $\Delta T(r) = T_{\rm atm}(r) - T_{\rm mid}(r)$ and we choose $z_q = 3$ and $\delta = 2$. The temperature and density structure for a column located at $r = 30$ au is shown in figure~\ref{fig:phys_structure}.

We also assume that the disk is exposed to an external UV radiation field.  As described above, the depth of UV penetration is controlled by the dust abundance.  The optical depth into the disk at each location in the column, $\tau_{UV}(z)$, is found via numerical integration:
\begin{equation}
    \label{eq:opacity}
    \tau_{UV}(z) = \int_z^\infty \epsilon(z') \kappa_d \rho_g(z') \text{ d}z'
\end{equation}

\noindent where $\epsilon(z)$ is the local small dust-to-gas mass ratio and the dust opacity is picked to be $\kappa_d = 10^5 \text{ cm}^2 \text{ g}^{-1}$ to agree with the extinction-column density relationship of \citet{2001A&A...371.1107A}.  Here, the UV optical depth is assumed to be dominated by extinction from the small dust, with the gas opacity $\kappa_g \approx 0$ such that $\kappa_{tot} \approx \epsilon \kappa_d$.

Likewise, we determine the column density of \ce{CO} above each given vertical height by integrating the abundance of the molecule relative to the number density of hydrogen $X(\ce{CO},z) = n_{\ce{CO}}(z)/n_H(z)$, where $n_H(z) = 2\rho_g(z)/\mu m_H$

\begin{equation}
    N_{\ce{CO}}(z) = \int_z^\infty X(\ce{CO},z')n_H(z')\text{ d}z'.
\end{equation}
A similar methodology is used to determine the column abundance of H$_{2}$ above a given location.  These column densities  are used to account for self-shielding by these molecules, as discussed in section 2.4.

While we allow the dust and UV vertical profile to evolve over time, the temperature and density structure remain fixed throughout our simulation, following \citet{2020ApJ...899..134K}. In future implementations of this model, we will include evolving density and temperature structures with the evolving UV radiation.

\subsection{Turbulence and Diffusion}
While the temperature and density structures remain constant throughout our simulations, we assume that materials are able to diffuse due to the presence of some background turbulence.  We define the diffusivity of the gas as $D = \alpha c_s H$, and take it to be constant throughout the slice of the disk, with $c_s$ evaluated at the midplane. The value of $\alpha$ in disks is uncertain; we assume a value of $\alpha$=10$^{-3}$ for our fiducial case, but also investigate other values as described in Section \ref{CO_depl}.  

 For a chemical species mixed within the gas and abundance, $X_i$,  the vertical transport of that species is given by:

\begin{equation}
\label{eq:diffusion}
    \frac{\partial X_i}{\partial t} = \frac{1}{\rho_g} \frac{\partial}{\partial z} \left(\rho_g D_i \frac{\partial X_i}{\partial z}\right)
\end{equation}
for all species considered \citep{2018ApJ...864...78K}.

Although settling of dust is not directly considered, dust (and ice) species are removed near the midplane in pebble growth. As a result, there is a net downward diffusive transport of dust and ice, effectively leading to dust settling in the column. The net effect is a reduction in fine dust in the upper layers and a concentration of solid mass in pebbles at the midplane.

\subsection{Dust Growth and Pebble Formation}

The solids in our model consist of two populations as in \citet{2012A&A...539A.148B}: small grains that are coupled to the gas and larger pebbles that decouple and are removed from our simulation after they form.  The small grains interact with the gas, serving as sites for freeze-out and chemistry as well as being a source of gaseous species as they desorb from grain surfaces.   The small grains are assumed to have radii $s = 0.1 \mu\rm{m}$ and are initially present at a dust-to-gas mass ratio, $\epsilon$, of $10^{-2}$ everywhere in the column. We set the initial abundance of grains, $X(\rm{grain}) = 2.2\times10^{-12}$ following \citet{2018AA...618A.182B}, setting a material density of $\rho_{\rm m} = 1.81 \rm{\ g\ cm^{-3}}$.  The small sizes and low densities imply that these grains are fully coupled to the gas and thus their abundance over time is described by Eq.~\ref{eq:diffusion}, with the same diffusivity.
 
Pebbles form from the small dust grains via collisional growth.  In detail, this would result from incremental growth of monomers and smaller aggregates  over time.  Rather than investigating this computationally intensive process here, we follow \citet{2020ApJ...899..134K} by defining a timescale for pebble formation motivated by previous detailed studies of the process \citep[e.g.][]{2012A&A...539A.148B}:

\begin{equation}
    \label{eq:growth_time}
    \tgrow = a\frac{1}{\Omega \epsilon}
\end{equation}
where $a$ is a scaling factor of order unity. This is used to define a loss of dust in a given cell to pebble formation that is described by:

\begin{equation}
\label{eq:deps}
    \frac{\partial \epsilon}{\partial t} = - \frac{\epsilon}{\tgrow}
\end{equation}
As growth via collisions will be most efficient where dust densities are highest, pebbles are largely expected to form around the disk midplane.  In our model we thus limit pebble formation to within one scale height of the disk midplane. We explore impacts of allowing pebbles to grow at all elevations in the disk in sections 4 and 5.

The pebbles that form at a given time reflect the composition of the dust that is present.  Thus we likewise adjust the abundance of any species that is frozen out onto the dust similarly to Eq.~\ref{eq:deps}:

\begin{equation}
\label{eq:dice}
    \frac{\partial X(\rm{ice})}{\partial t} = -\frac{X(\rm{ice})}{\tgrow}.
\end{equation}

In this treatment, we assume that once pebbles form, their ices are locked away and removed from interacting with the gas and dust in the column.  This assumption is valid provided that pebble formation is limited by the drift or bouncing barriers, rather than fragmentation barriers, appropriate for the outer regions of a protoplanetary disk considered here \citep{2012A&A...539A.148B,2019ApJ...885..118M}

\subsection{Chemical model}

Chemical evolution is modeled using a modified version of the publicly available \texttt{astrochem} solver \citep{2015ascl.soft07010M}.
For the gas phase reactions, we use reaction rate coefficients from the UMIST 2012 database \citep{2013A&A...550A..36M}.
Self shielding of CO and H$_2$ were also included following \citet{2009A&A...503..323V}.  Additionally, we also account for chemistry with excited \ce{H_2} molecules \citep[Appendix A]{2018A&A...615A..75V}.

Thermal and photodesportion rates are determined following \citet{2011A&A...534A.132V}. This limits the
desorption rate by ensuring that each species
desorbs as a function of its solid-phase abundance in the ice mantle.
Binding energies are also taken from the UMIST 2012 database 
\citep{2013A&A...550A..36M}.
In addition, hydrogenation reactions were added to the chemical network also following  \citet{2011A&A...534A.132V}.
Reactions with hydrogen dominate gas grain chemistry on short timescales \citep{hasegawa1992models} and as such, hydrogenation of atomic species considered in \citet{2011A&A...534A.132V} is the only gas-grain process considered here beyond freezeout and desorption. A list of hydrogenation reactions considered is given in table~\ref{tab:hydrogenation} in the appendix. As a result, CO does not react with other species when present as an ice.

These modifications make the gas and grain chemistry more similar to the chemical calculations of the DALI code \citep{2012A&A...541A..91B}. 
A detailed description of our network and chemical model is given in the appendix section \ref{sec:chem_model}.
Our complete network includes 6830 reactions including 741 species.

\subsection{Numerical Model}

For our model, we define a vertical column from the midplane ($z=0$) to 5 scale heights ($z=5H$) at a given stellar distance, $R$. The vertical column is divided into 50 cells, each with a height $dz = 0.1 H$. Initially each cell has molecular  abundances modified from \citet{2018AA...618A.182B} given in Table~\ref{tab:init_abuns}. For each cell, the initial density, temperature, column densities, optical depth, and dust-to-gas ratio are calculated.

\begin{deluxetable}{lr|lr}
    \tablehead{
        molecule & abundance &
        molecule & abundance
    }
    \startdata
        \ce{H2} & 5.00(-1) & \ce{He} & 9.75(-2) \\
        \ce{NH_3} & 1.45(-6) & \ce{H_2O} & 1.18(-4) \\
        \ce{CO} & 6.00(-5) & \ce{N_2} & 2.00(-5) \\
        \ce{CH_4} & 2.00(-6) & \ce{CH_3OH} & 1.00(-6) \\
        \ce{H_2S} & 1.91(-8) & \ce{CO_2} & 5.00(-5) \\
        \ce{HCN} & 3.50(-7) & \text{grains} & 2.20(-12) \\
    \enddata
    \caption{Initial abundances in each cell. Modified from \citet{2018AA...618A.182B}.}
    \label{tab:init_abuns}
\end{deluxetable}

\begin{figure}
    \centering
    \includegraphics[width=0.98\linewidth]{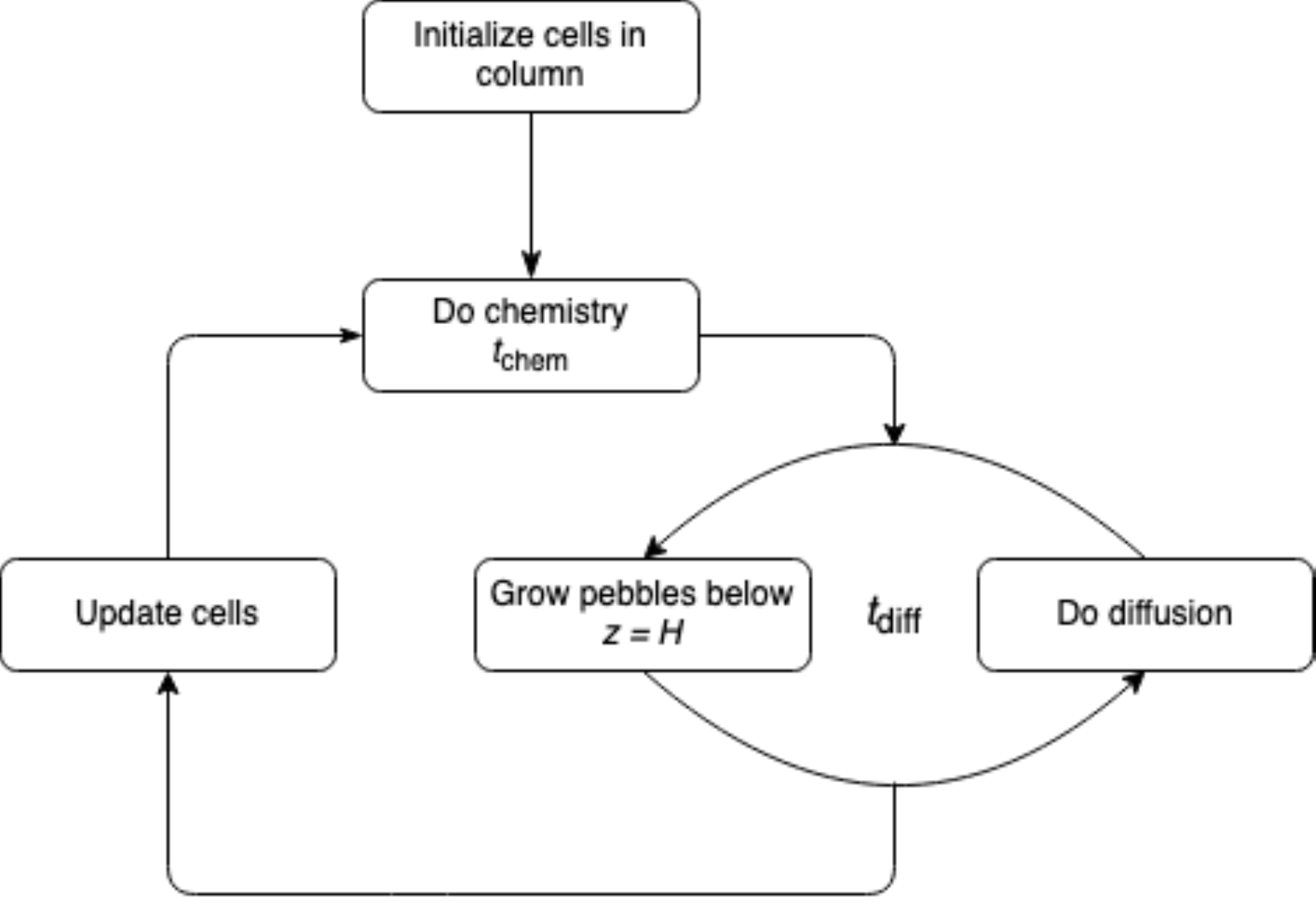}
    \caption{A schematic diagram of the process of the model presented in this paper. We run our models for a total of $t_{\rm tot} = 1\text{ Myr}$ with $t_{\rm chem}=500 \text{ yr}$ and $t_{\rm diff}=100 \text{ yr}$. In other words, the chemical model is run a total of 2000 times (m=2000) for each cell, and the diffusion calculation is done 5 times (n=5) between each chemical model.}
    \label{fig:model_schematic}
\end{figure}

The subsequent calculations are then carried out as illustrated in Figure~\ref{fig:model_schematic}.
We begin by running our modified \texttt{astrochem} model on each cell for a time, $t_{\rm{chem}}$ where $t_{\rm{chem}} < t_{\rm tot}$ and $t_{\rm{tot}}$ is the total simulation time. Once the chemical calculations are completed, the abundances of each species are updated.  We then calculate the diffusion of each individual species through the disk over a diffusion time step, $t_{\rm diff}$, using an Euler approach and ensuring that:

\begin{equation}
    t_{\rm{diff}} < \frac{\left( 0.1 H \right)^{2}}{D} = \frac{1}{100} \left( \alpha \Omega_{K} \right)^{-1}
\end{equation}

Once species abundances are updated via diffusion, pebble growth occurs, reducing the abundances of frozen-out species within one scale height of the disk midplane following Eqn.~\ref{eq:dice} for the same length of time, $t_{\rm{diff}}$.  The diffusion and growth calculations are repeated again until a total time interval of $t_{\rm{chem}}$ was reached.   While setting $t_{\rm{chem}}$=$t_{\rm{diff}}$ would be ideal, it can be computationally expensive; we find  $t_{\rm chem} = n t_{\rm{diff}}$ with $n$=5 yields convergent results while still giving reasonable run times.
 
 After repeating the diffusion and dust growth step $n$=5 times, new column densities, optical depth, and dust-to-gas ratios are determined for each cell. The chemistry is then run again at each location and the process repeats until the cumulative time reaches $t_{\rm tot}$.

\section{Fiducial Cases}\label{fiducial}

To begin, we consider the chemical evolution that occurs in our model disk at 30 and 40 au from the central star, regions that are just inside and outside of the CO snow line respectively.  We pick these locations as we expect significant differences in carbon chemistry as a result of freeze-out of CO at the outer location.  In order to isolate the consequences of pebble growth on the chemistry that occurs, we first consider two simpler disk dynamic scenarios: a \emph{static} model and a \emph{diffusive} model.  Static modeling is the simplest and most common approach used in disk chemistry studies, in which the chemistry evolves at all heights of the disk under fixed conditions
\citep[e.g.][]{2012A&A...541A..91B, 2015ApJ...807L..32D, 2015A&A...579A..82R, 2015A&A...582A..88W, 2017A&A...607A..41K, 2018ApJ...865..155C, 2018ApJ...856...85S, 2021arXiv210906286L, 2021arXiv210906221B}.  In a diffusive model, chemistry occurs while all materials are simultaneously able to diffuse through the column \citep[e.g.][]{2011ApJS..196...25S, 2016A&A...592A..83K}. We present these models to compare with our \emph{growth} model, which includes both pebble growth and diffusion as outlined above. The parameters used in these fiducial models are  given in table~\ref{tab:fiducial_conditions}.  

\subsection{Static and Diffusive models}

\begin{deluxetable}{lcl}
    \tablehead{
        Parameter & Symbol &
        Value
    }
    \startdata
        UV Flux & $\chi$ & 50 \\
        Cosmic-ray ionization rate & $\zeta$ & $1.3\times10^{-17}$ s$^{-1}$ \\
        Disk turbulence parameter & $\alpha$ & 10$^{-3}$ \\
        Dust growth timescale parameter & $a$ & 1 \\
    \enddata
    \caption{Fiducial model parameters. UV fluxes are given in Draine units \citep{1978ApJS...36..595D}.}
    \label{tab:fiducial_conditions}
\end{deluxetable}

\begin{figure*}
    \centering
    \includegraphics[width=0.8\linewidth]{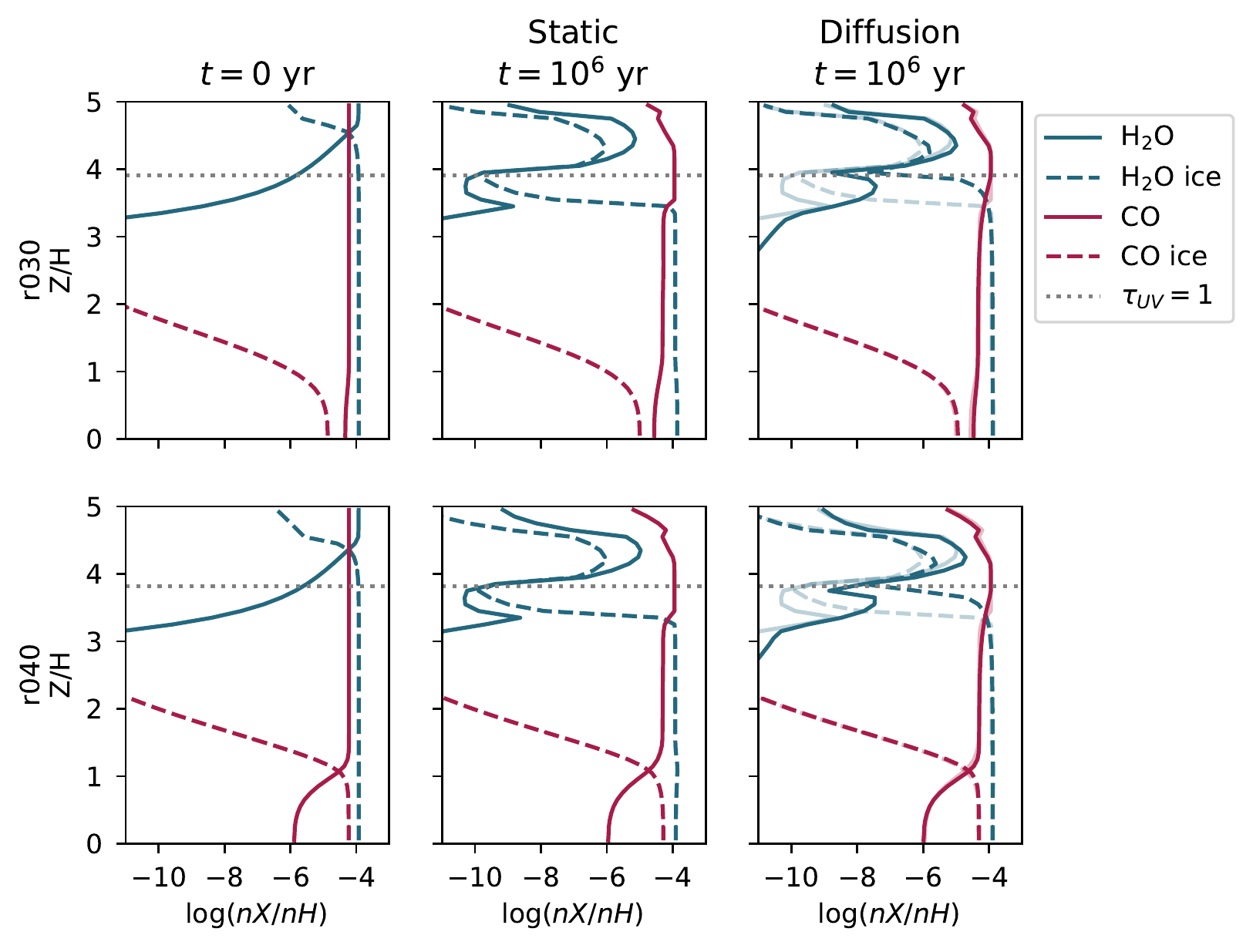}
    \caption{Comparison of initial conditions (left) inside (30 au, top) and outside (40 au, bottom) the CO snowline. Results after 1 Myr evolution for the static (middle) and diffusive (right) are also shown. The static evolution model is also shown as transparent lines in the final panel to help compare results.}
    \label{fig:compare_static_and_diffusion}
\end{figure*}

Figure~\ref{fig:compare_static_and_diffusion} shows the distribution of CO and H$_{2}$O in the static and diffusive models.  We focus on these two species to explore the differences in the behavior of key molecules with different volatilities.  The initial distributions of each species in the ice and gas are also shown for reference; these are found by balancing thermal and photo-desorption with adsorption onto the grains in the absence of any other chemistry.  At the locations considered here, temperatures are low enough that water is initially frozen out onto dust grains throughout most of the column, with the exception of the upper ~0.5 scale heights of the disk where photodesorption is efficient. At 30 au, CO is entirely in the gas phase except for a very minor amount of CO ice that forms on the grains. At 40 au, CO is nearly completely frozen out as an ice around the midplane, but as temperatures rise at higher elevations, it returns to the gas phase. 

 Within the static models, we see a significant drop in the water abundance above the  $\tau_{UV} = 1$ surface (which we refer to as the ``photon dominated layer'' or ``PDL'' in subsequent discussion).  This is due to the high flux of UV, which leads to photodesorption off of the grains and photodissociation in the gas.  Water is not completely destroyed in this region, however, as water can reform via reactions involving vibrationally excited H$_{2}$ and gas phase nitrogen. \citep[see][appendix A]{2018A&A...615A..75V}.  Deeper in the disk, below the photon dominated layer, water largely remains on the ice due to the low temperatures and low photon fluxes present.
 
Similarly, the CO abundance and distribution between the gas and solid phases remains largely unaffected below the photon dominated layer.  Above this, CO is destroyed by photodissociation but also produced by the photodesorption of CO$_2$ and its subsequent chemical processing.  The net result is a slight increase in the CO abundance around the PDL.

When diffusion is considered, we see only slight changes in the H$_{2}$O and CO distributions when compared to the static cases.  The net effect of diffusion is to smooth out concentration gradients.  This is clearly seen in the H$_{2}$O distribution around the $\tau$=1 line, where diffusion carries water-ice bearing grains to higher altitudes than in the static model.  This occurs because the chemical timescales are finite, rather than instantaneous. For example, around the PDL the timescale for removing all the water from the grains is on the order $\sim$1000 years, which is longer than the residence time of grains in this region. As a result, ice-covered grains can move to higher altitudes compared to the static model.  Where there is little difference between the diffusive and static cases, the timescales for reactions are short compared to the residence times.

\subsection{Pebble growth model}

\begin{figure*}
    \centering
    \includegraphics[width=\linewidth]{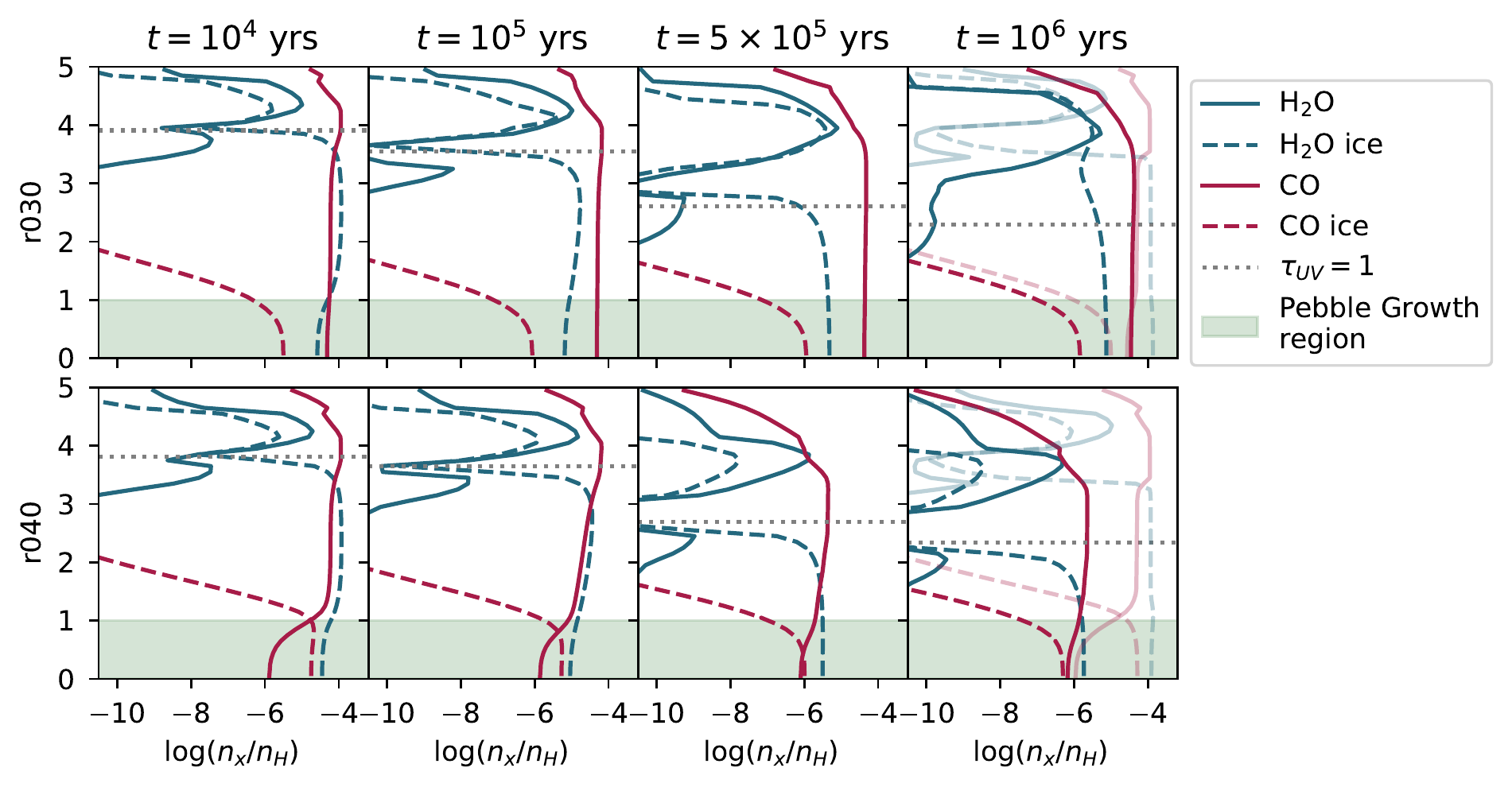}
    \caption{Comparison of the time evolution of CO and H2O abundances inside (top) and outside (bottom) the CO snowline for growth models including pebble growth and diffusion. Static abundances are shown as transparent lines in the rightmost panel as a comparison.}
    \label{fig:compare_evolution}
\end{figure*}

Fig.~\ref{fig:compare_evolution} shows the time evolution of CO and H$_{2}$O distributions in our pebble growth models for the same locations considered for the static and diffusive models previously. 

As dust is removed, pebble formation acts as a physical sink for icy molecules around the disk midplane.  Here, any species that are frozen out onto dust grains are sequestered, removing them from the active chemistry and dynamical redistribution occurring in the disk. As ice and grains are removed from the midplane, diffusion acts to transport material downward through the column, replenishing the midplane in new ices, and depleting the upper layers of the disk, even though pebbles are not forming there.  In the case of H$_{2}$O, this results in its total abundance dropping by 1 to 2 orders of magnitude throughout the column at both 30 and 40 au.

Similarly, diffusion and pebble growth deplete the fine dust abundance in the upper layers of the disk. As the fine dust is the major source of UV opacity, 
the photon dominated region migrates deeper inside the disk, much closer to the midplane.
The increased extent of the PDL drives photochemistry in denser regions of the disk, leading to the increased destruction of CO and chemical processing via excited H$_{2}$ chemistry compared to the static or diffusive models.

\begin{figure}
    \centering
    \includegraphics[width=\linewidth]{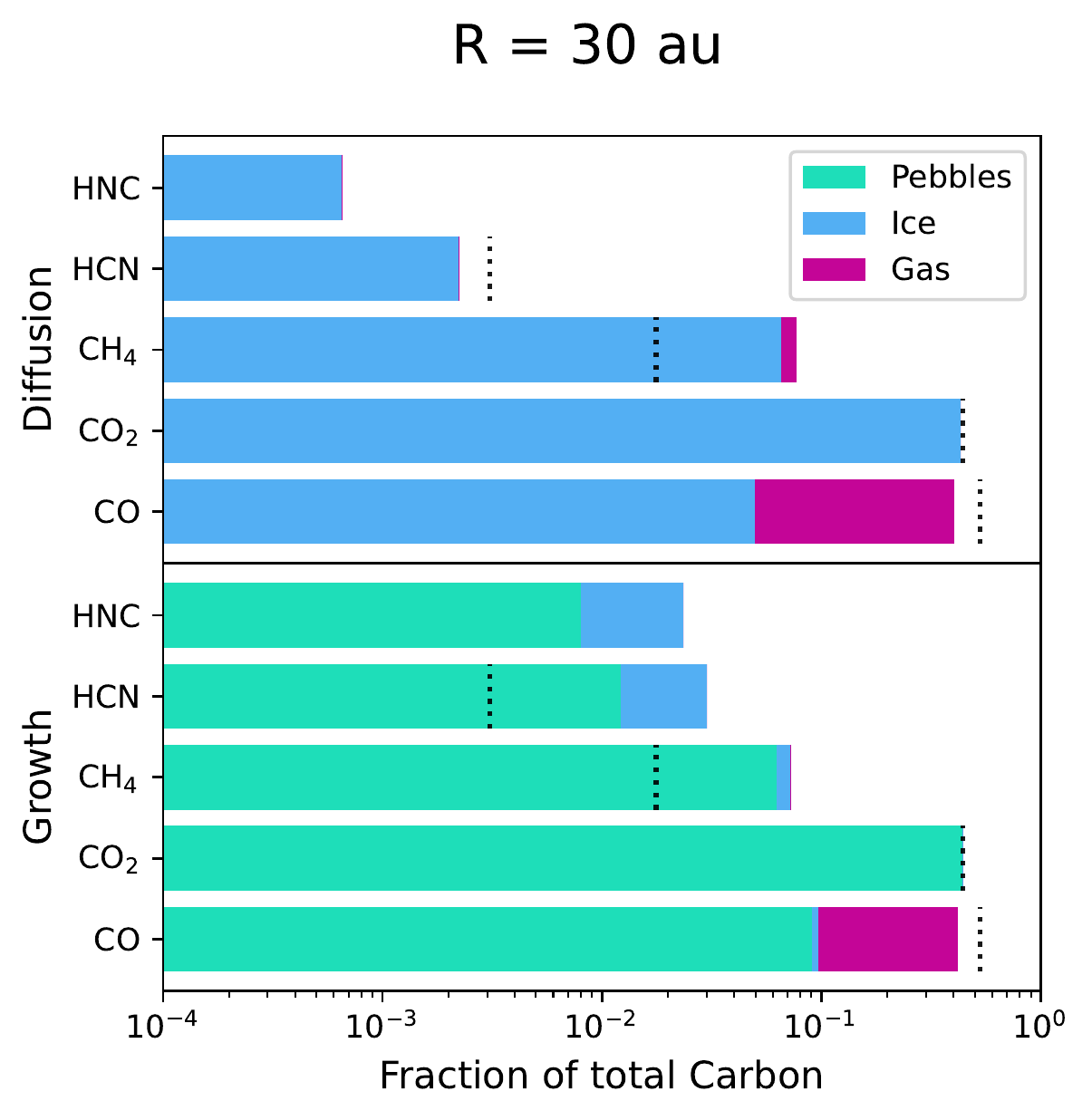}
    \caption{Fraction of carbon present in secondary species for the diffusive (top) and pebble growth models (bottom) at 30 au after 1 Myr. The distribution of each species into the gas, ice on small dust, and ice in pebbles is also shown. For species where the initial abundance was not zero, the black dotted line shows the initial carbon fraction. Compared to the diffusive model, about an order of magnitude more carbon is incorporated into the photoproducts HCN and HNC in the growth model. This is due to increased UV-driven chemistry in the disk as a result of the removal of small dust.}
    \label{fig:r030_carbon_products}
\end{figure}

\begin{figure}
    \centering
    \includegraphics[width=\linewidth]{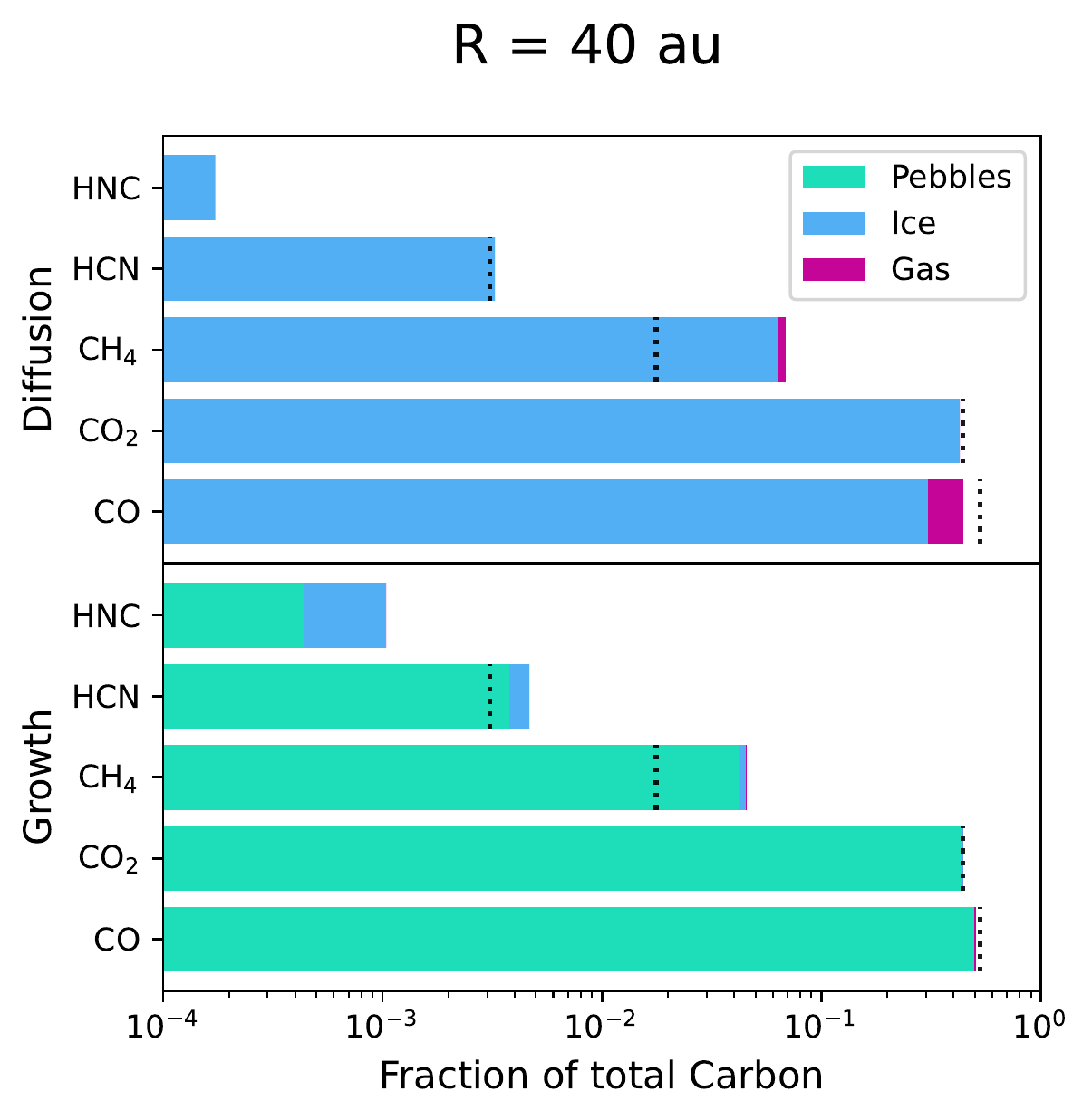}
    \caption{Same as figure~\ref{fig:r030_carbon_products} but at 40 au, outside the CO snowline. Relative fractions of all secondary carbon carriers is lower than in the 30 au case, as CO is primarily in the ice phase, and thus is not as readily processed.}
    \label{fig:r040_carbon_products}
\end{figure}

Figures~\ref{fig:r030_carbon_products} and \ref{fig:r040_carbon_products} show the primary carbon carriers, and their distribution across the different reservoirs of carbon. Species in the gas and ice reservoirs are active in chemistry, and represent the gas phase or ice frozen to small dust grains, while the pebble reservoir contains the ice that has been sequestered into larger pebbles. For species that were present initially, the black dotted line shows the initial fraction of the total carbon.
Both inside and outside the CO snowline, dust depletion and increased UV-radiation in the growth model increase production of HCN and HNC compared to the diffusive model. As dust is removed, increased UV-radiation excites H$_2$ in denser regions of the disk. This excited H$_2$ can react with atomic N (dissociated from N$_2$, again due to higher UV radiation) to create NH, in an otherwise endothermic reaction \citep{2018A&A...615A..75V}. The NH molecule can then react with atomic C, produced from the dissociation of CO, to create the cyanides HCN and HNC.

The carbon that winds up in HNC, HCN, and other secondary carriers is largely sourced from the  destruction of gas phase CO. Given their lower volatilities relative to CO, this processing increases the carbon sink that occurs via pebble formation. At 40 au, CO is largely removed via pebbles directly, limiting the amount of secondary materials produced when compared to 30 au.

The net effect of pebble growth on the CO distribution is very different at the two locations considered here. At 30 au, CO remains largely in the gas phase even close to the midplane, so pebble formation does not significantly contribute as a sink for the molecule.  As a result, CO is primarily destroyed by photoprocessing in the PDL.  It is important to note that this does not occur throughout the entirety of the PDL, as CO self-shielding limits CO dissociation to the upper-most layers of the disk.  The story is different at 40 au, however, as CO is frozen out at the disk midplane, and thus removed from the column via pebble formation.   Diffusion from upper layers of the disk continuously replenish the midplane region with additional CO to be sequestered into pebbles. This effect, combined with the increased penetration of UV in the disk (due to removal of both dust and CO) leads to a decrease in CO throughout the column by at least 2 orders of magnitude after 1 Myr, with the greatest levels of depletion at higher altitudes (see fig~\ref{fig:compare_evolution}).

Thus, the interplay of chemistry, diffusion, and pebble formation is critically important in shaping the distribution of molecules throughout a column of the disk.  In the next sections we explore how this interplay depends on the conditions present in the disk by focusing on two important themes related to protoplanetary disk observations: CO depletion and the C/O ratio.

\section{CO Depletion During Pebble Growth}
\label{CO_depl}

\begin{figure*}
    \centering
    \includegraphics[width=\linewidth]{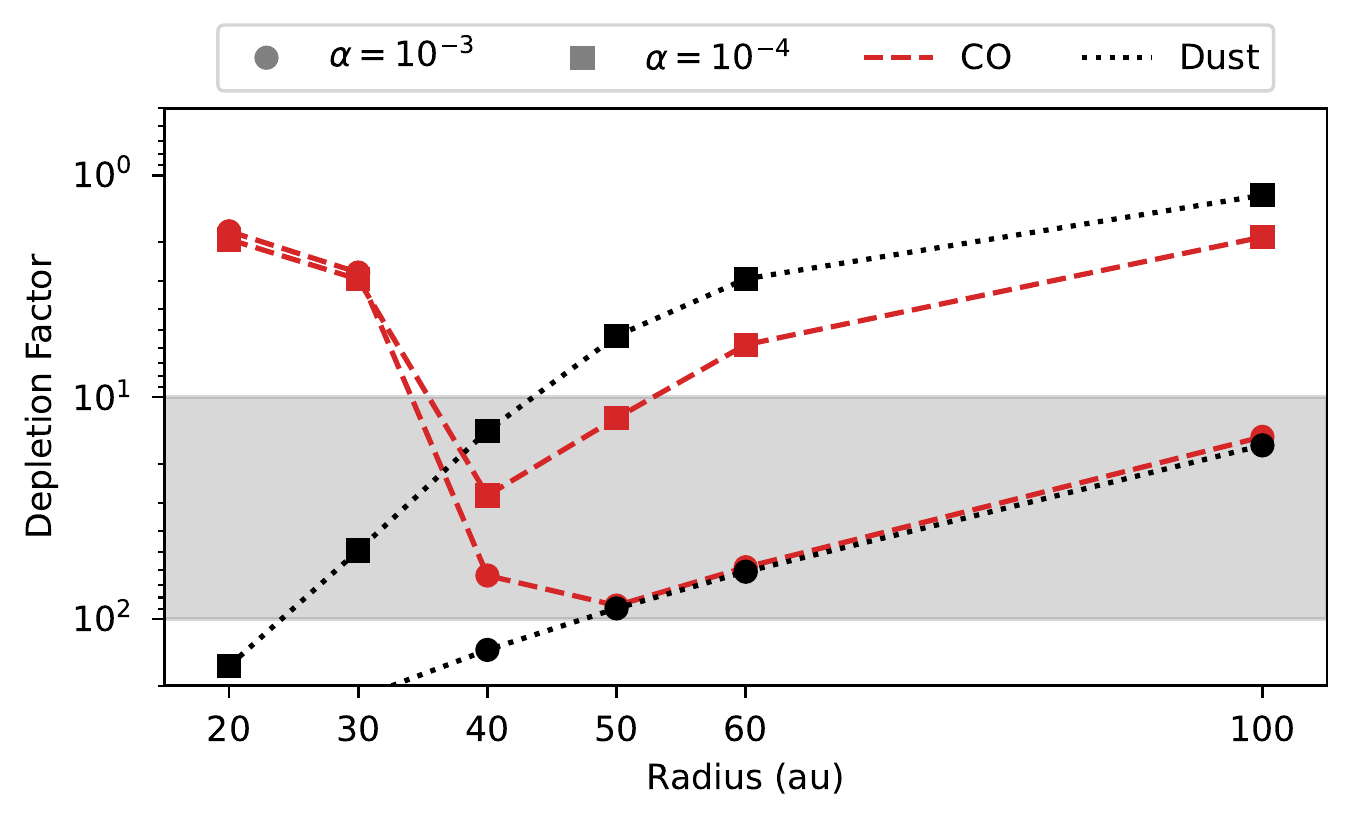}
    \caption{CO and dust depletion relative to ISM abundance of $10^{-4}$ and small dust-to-gas ratio of 0.01 in the warm molecular region ($T>21$K) as a function of radial distance after 1 Myr of evolution. CO abundance generally follows the evolution of small dust, shown in black, outside of the CO snowline ($\sim$35 AU). Observed CO depletion ranges of 1-2 orders of magnitude are shaded in grey.}
    \label{fig:co_depletion}
\end{figure*}

As discussed in the previous section, the depletion of CO in a column will arise due to two active sinks: chemical processing of CO gas that largely occurs in the PDL, and pebble formation around the midplane that sequesters CO ice into large dust aggregates.  The extent and time over which these processes operate will determine the amount to which CO is depleted in the column and timescale over which this depletion occurs.  Figure~\ref{fig:co_depletion} shows the CO depletion within the emissive layers of the disk, defined as T $> 21$ K \citep{2020ApJ...899..134K},  after 1 Myr at various radial locations in the disk. The depletion factor is given by

\begin{equation}
    f_{\text{depl}} = \frac{X_0}{X_f}
\end{equation}

\noindent where $X_f$ is the abundance of CO or small dust grains after 1 Myr evolution, and $X_0$ is the initial abundance: $10^{-4}$ for CO and $2.2\times10^{-12}$ for small dust (small dust-to-gas ratio of $\epsilon=0.01$).
In each panel, different sets of parameters are considered.  We compare the modeled depletion levels to observed CO depletion factors, which are on the order of 1-2 orders of magnitude below interstellar \cite{2013ApJ...776L..38F, 2019ApJ...883...98Z}. Notably, CO depletion is a robust outcome independent of our model assumptions, though the extent of depletion is sensitive to the different conditions present within the disk.

For the fiducial model, we find that the greatest CO depletion occurs in the region just beyond the CO snow line (40-60 au), while the CO abundance rises again at larger distances (100 au).  This indicates that the formation of pebbles and sequestration of ices plays a critical role in CO depletion; if photoprocessing alone were more important, we would expect higher depletion levels in the outer disk where UV can penetrate more deeply.  The lack of CO depletion inside the CO snow line is further evidence of the importance of pebble formation in reducing CO levels in the disk.

The higher depletion levels around the CO snowline compared to the outer disk is due to the shorter time scales associated with dust growth and diffusion of small gas. As the growth time scale of pebbles scales with orbital period (see Eq.~\ref{eq:growth_time}), a greater fraction of dust is converted into pebbles within 1 Myr at smaller radii, corresponding to more efficient removal of CO ice.  The slightly lower depletion of CO at 40 au instead of 50 au in some cases is due to the shallow temperature gradient in the disk: there is a non-trivial amount of CO that remains in the gas phase at the disk midplane at 40 au despite the ice being the dominant form there. At larger distances, the amount of gas phase CO at the midplane is negligible.

Diffusion plays an important role in setting the level of CO depletion that occurs in the disk.  The effect of different $\alpha$ values on the extent of CO depletion after 1 Myr is shown in the top panel of figure~\ref{fig:co_depletion}.  In our model, pebble formation only occurs around the disk midplane; diffusion of dust downward reintroduces solids and allows ices to be further sequestered in the model.  For pebble formation to be an effective sink for the CO, diffusion is necessary to carry the CO to greater depths where it can freeze out as an ice. Thus, the extent and timing of depletion is intimately connected to the level of turbulence present. The timescale for vertical diffusion  is $t_{diff}$ = $H^{2}$/$D$ = $\left( \alpha \Omega \right)^{-1}$.  At 40 au, this gives $t_{diff} \sim 4\times$10$^{4}$ years for $\alpha$=10$^{-3}$, and 10$\times$ greater for $\alpha$=10$^{-4}$.  
Faster diffusion results in more efficient removal of CO when there is an active sink around the midplane.  Given that significant CO depletion is seen across a wide variety of disks, including young disks, this points to an active environment that enables efficient vertical transport of gases in disks.

The relative roles of dust dynamics and chemical processing in shaping the CO distribution can be seen when comparing the level of CO depletion to the level of dust depletion in Fig.~\ref{fig:co_depletion} (top).  In the $\alpha$=10$^{-3}$ case, the level of CO and dust depletion are nearly identical at 50 AU or beyond, indicating that CO depletion is most strongly controlled by pebble formation.  Closer in, because much of the CO remains in the gas phase, even around the midplane, the dust is depleted to a greater level than the CO.  However, in the $\alpha$=10$^{-4}$ cases, the dust is depleted to a lesser extent than the CO.  This indicates that chemistry and photoprocessing contribute more to CO depletion than in the more turbulent cases. 

The extent of CO depletion in our models depends somewhat on the physical parameters of the disk. The details of these differences are discussed in the appendix \ref{co param sensitivity}. However, regardless of physical parameters, CO depletion is a robust outcome of our model, although to match observed depletions of 1-2 orders of magnitude high turbulence ($\alpha = 10^{-3}$) is necessary.

\section{C/O Response to Ice Sequestration}

As discussed in the previous section, pebble formation aids in the depletion of volatiles in a disk by the removal of ices around the midplane.  The removal of ice species may impact subsequent gas-phase chemistry by changing the abundances and proportions of various reactants.  For example, \citet{2016ApJ...831..101B} suggested that preferential removal of O-bearing species on pebbles may produce high C/O ratios, leading to the formation of hydrocarbons such as C$_{2}$H.  Indeed, C$_2$H formation is favored by C/O values $>$ 1.5-2, approximately 4-5$\times$ the canonical value to be inherited from the ISM \citep{2021ApJ...910....3B}.

\begin{figure}
    \centering
    \includegraphics[width=\linewidth]{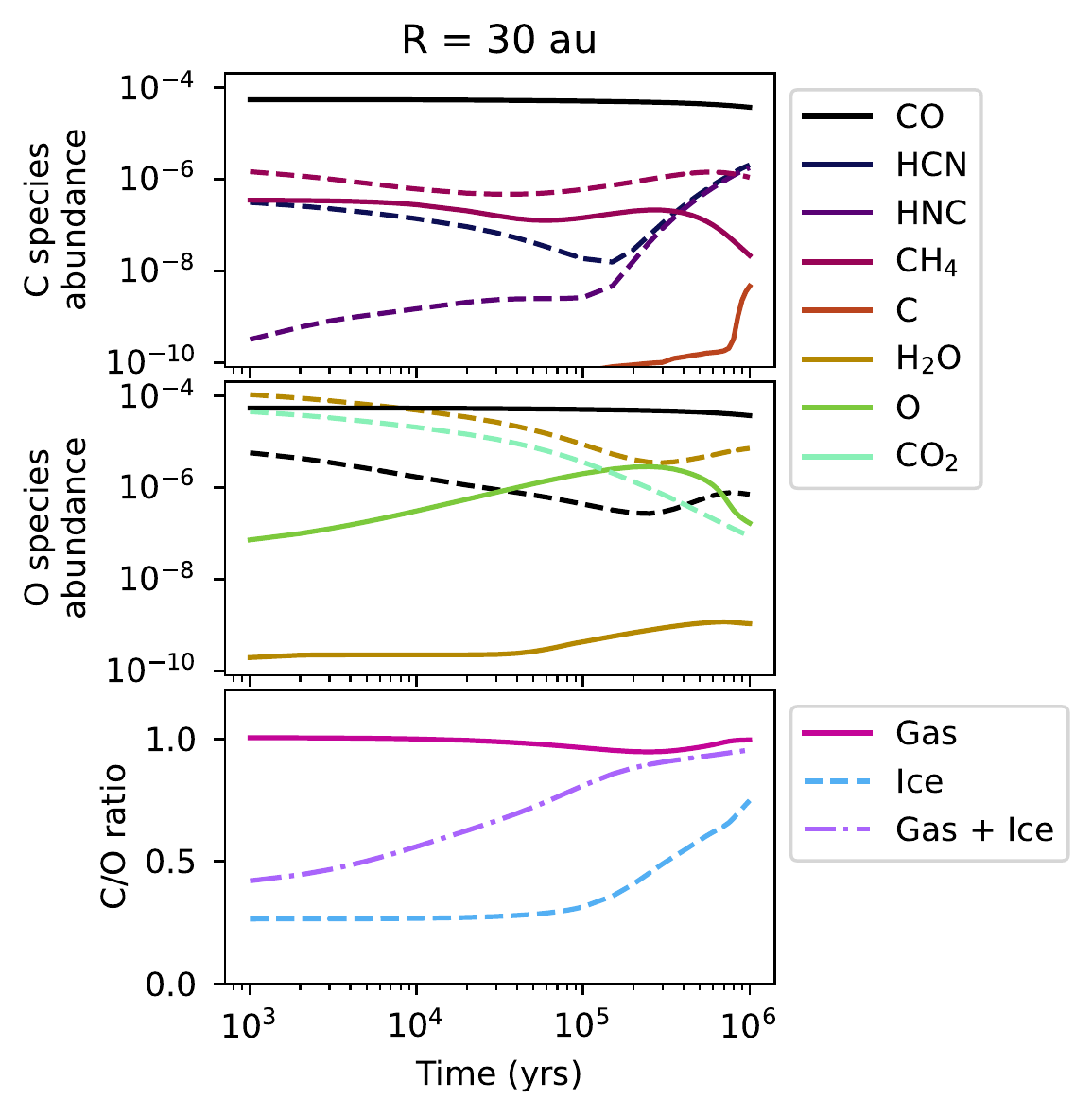}
    \caption{Most abundant column averaged carbon (top) and oxygen (middle) bearing species in the gas and ice. Dashed lines indicate ice phase abundances and solid lines represent gas phase abundances. Bulk C/O of the gas, ice, and ice and gas combined (bottom)}
    \label{fig:CO_and_specs}
\end{figure}

Figure~\ref{fig:CO_and_specs} shows the evolution of the major carbon- and oxygen-bearing species at 30 au along with the bulk C/O ratio in ice and gas.  While water is initially the most abundant oxygen carrier, its abundance drops rapidly due to the formation and removal of pebbles. The next most important oxygen bearing species from our starting inventory is CO$_{2}$; while CO is more abundant, CO$_{2}$ carries twice the amount of oxygen (see table~\ref{tab:init_abuns}).  At 30 au, CO$_{2}$ is largely present as an ice and is removed from the column along with water.  The loss of H$_2$O and CO$_2$ ice leads to a net increase in the  C/O ratio in the remaining gas + ice.  As pebble growth removes ice, the total gas + ice C/O ratio becomes weighted more heavily towards the gas C/O, which is now dominated by CO, leading to C/O$\sim 1$. As oxygen carriers become less and less abundant, the ices become increasingly rich in carbon and nitrogen, with cyanides the dominant carbon carriers in the ice by 1 Myr.  

\begin{figure}
    \centering
    \includegraphics[width=\linewidth]{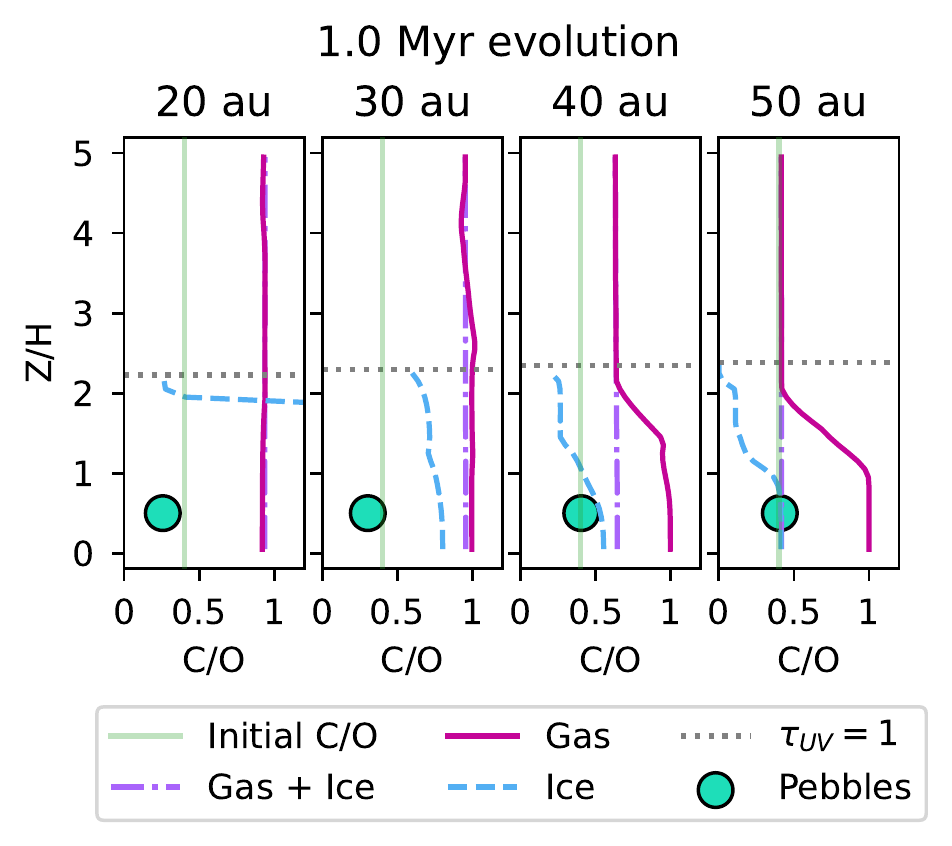}
    \caption{C/O ratios of the disk profiles at 20 au (left), 30 au (middle left), 40 au (middle right) and 50 au (right). Locations inside the CO snowline (30 and 40 au) tend to have C/O ratios close to unity throughout the column, while regions outside the C/O snowline (40 and 50 au) do not have C/O ratios above the inherited value. The initial C/O of 0.4 is shown by the green vertical bar.}
    \label{fig:CO_profile_1myr}
\end{figure}

The evolution of the C/O ratio in the column will thus be sensitive to the partitioning of C and O into carriers with different volatilities, which will vary strongly with location.  This is illustrated in
Figure~\ref{fig:CO_profile_1myr} which shows the C/O ratio of the gas and ice, as well as the total (gas+ice) C/O ratio throughout the column after 1 Myr of evolution at various locations in the disk.  With the starting abundances in our model,the column begins with a C/O ratio of $\sim$0.4 throughout.  Inside of the CO snow line, the C/O ratio increases everywhere as O-rich ices, notably H$_{2}$O and CO$_{2}$ are removed via pebble growth as described above, but CO gas remains behind.  This allows C-rich ices, such as HCN, HNC, and CH$_4$, to develop later in disk evolution.  While this creates ice C/O ratios $>1$ in some regions of the disk, the ice abundances in these regions tend to be low, and the total ice+gas C/O ratios closely follow the gas C/O. Outside of the CO snow line, CO, CO$_{2}$, and H$_{2}$O are all removed equally, resulting in a negligible net effect on the bulk C/O ratio in the column.  C/O ratios are higher in the gas around the disk midplane, but this is due to the small amount of CO left in the vapor phase.

In none of our cases, do we see C/O ratios that rise above $\sim$1, reaching the levels that have been inferred from disk observations of hydrocarbons.  The key to reaching high C/O ratios is for oxygen to be incorporated into ices while carbon remains in the gas phase.  The loss of H$_{2}$O and CO$_{2}$ via pebble growth produces C/O ratios of $\sim$1 within the CO snow line.  Going beyond this value is a challenge, and seemingly can only be achieved if CO is dissociated and the carbon is partitioned into a more volatile carrier than H$_2$O. However, as discussed in section 3.2, the majority of secondary carbon species are present as ices, and so both the carbon and oxygen from any dissociated CO is removed from the gas phase, leaving the total C/O unaffected.

This result is robust across the physical conditions considered, though the details are discussed in section \ref{C/O param sensitivity}. In general, the chemical processing of gas phase CO is too slow to raise C/O ratios above unity inside the CO snowline despite the removal of the oxygen rich ice. Outside of the CO snowline, carbon and oxygen are removed equally and there is no net change to the C/O ratio in the column.

\section{Discussion}\label{discussion}

The growth of small dust into larger pebbles has two main effects on the gas phase chemistry of the disk
\begin{enumerate}
    \item removal of small dust from upper layers of the disk extends the photon dominated layer, and
    \item selective sequestration of ices can affect elemental abundances in the gas phase.
\end{enumerate}

These two effects occur simultaneously with the ongoing chemical evolution in the gas.  Collectively, this leads to important feedbacks, altering the chemistry that occurs over million year timescales compared to static or diffusive disk models.  Here we review our main findings and discuss how they can be used to interpret ongoing observations of protoplanetary disks.

\subsection{CO Depletion}

We find that CO depletion is ubiquitous outside of the snowline.  For our fiducial model we find depletion factors from $\sim$10-100 in the outer disk, with comparable or lower depletion factors (2--100) seen for different assumptions of turbulence, growth timescale, or dust formation height. This result is particularly important in light of the wealth of observational evidence that gas-phase CO is underabundant by up to 2 orders of magnitude in protoplanetary disks compared to the interstellar medium, on timescales of only $\sim$1 Myr \citep{2020ApJ...898...97B, 2020ApJ...891L..17Z}.

This finding expands upon
previous studies, which showed that CO can be depleted up to a factor of 10 in static chemical models \citep{2015A&A...579A..82R, 2018AA...618A.182B, 2018ApJ...856...85S}, or by a factor of $\sim$10 by freeze-out onto dust and pebble formation \citep{2016A&A...592A..83K, 2018ApJ...864...78K}.  As these studies treated either just chemistry or dust growth, it was suggested that the combined effects of both may be necessary to explain the observed depletions in disks.
\citet{2020ApJ...899..134K} was the first to combine physical and chemical processes, by including a simplified carbon chemical network in a pebble growth model.  They found that CO could be depleted by 2 orders of magnitude on a 3 Myr timescale in this framework.  Our model extends the chemistry beyond 5 carbon-bearing species and considers a full astrochemical network.  With this update, our model can achieve 2 orders of magnitude depletion of  CO  on a faster timescale ($\sim$1 Myr).  This is likely due, in part, to the inclusion of photochemistry in our model, which results in CO destruction in the disk surface layers in addition to ice sequestration in the midplane. 
Although our model does not include hydrogenation of CO on the grains, further processing of CO ice would not greatly affect the gas abundance of CO in our vertical diffusion model on 1~Myr timescales. In our model, gas phase CO is quickly sequestered into the pebbles, and so the growth mechanism presented here is the dominant source of CO depletion on shorter timescales ($\leq1$~Myr). However, this secondary processing of CO ice 
would further reduce the CO gas abundance on longer timescales and
may be more important in inner regions of the disk (inside the CO snowline) and when considering the inward drift of pebbles.
Our model can thus explain observations of CO depletion even in young ($\lesssim$1 Myr) disks \citep{2018ApJ...865..155C, 2020ApJ...898...97B, 2020ApJ...899..134K}. 

We also explored various physical parameters to test the robustness of these results.  Interestingly, we find that high values of $\alpha$ (10$^{-3}$) are required to explain CO depletion factors of 10--100 on 1 Myr timescales throughout much of the disk. 
Models with $\alpha = 10^{-4}$ show less depletion, around $\sim$ 30$\times$ immediately outside the CO snowline, and even lower (2$\times$) in the outer disk.  This is because diffusion is necessary to carry CO gas from the surface regions of the disk to the deeper interiors where it can freeze-out and be incorporated into growing pebbles.  This helps explain why \citet{2020ApJ...899..134K} were able to reach similar levels of depletion, but on longer timescales,as they used $\alpha$=10$^{-4}$.
 $\alpha$ values around 10$^{-3}$ are consistent with, if on the high end, of the range inferred from observations of mature protoplanetary disks \citep{2017ApJ...843..150F, 2018ApJ...869L..46D, 2018ApJ...864..133T}.  

\subsection{C/O ratios}

\begin{figure}
    \centering
    \includegraphics[width=\linewidth]{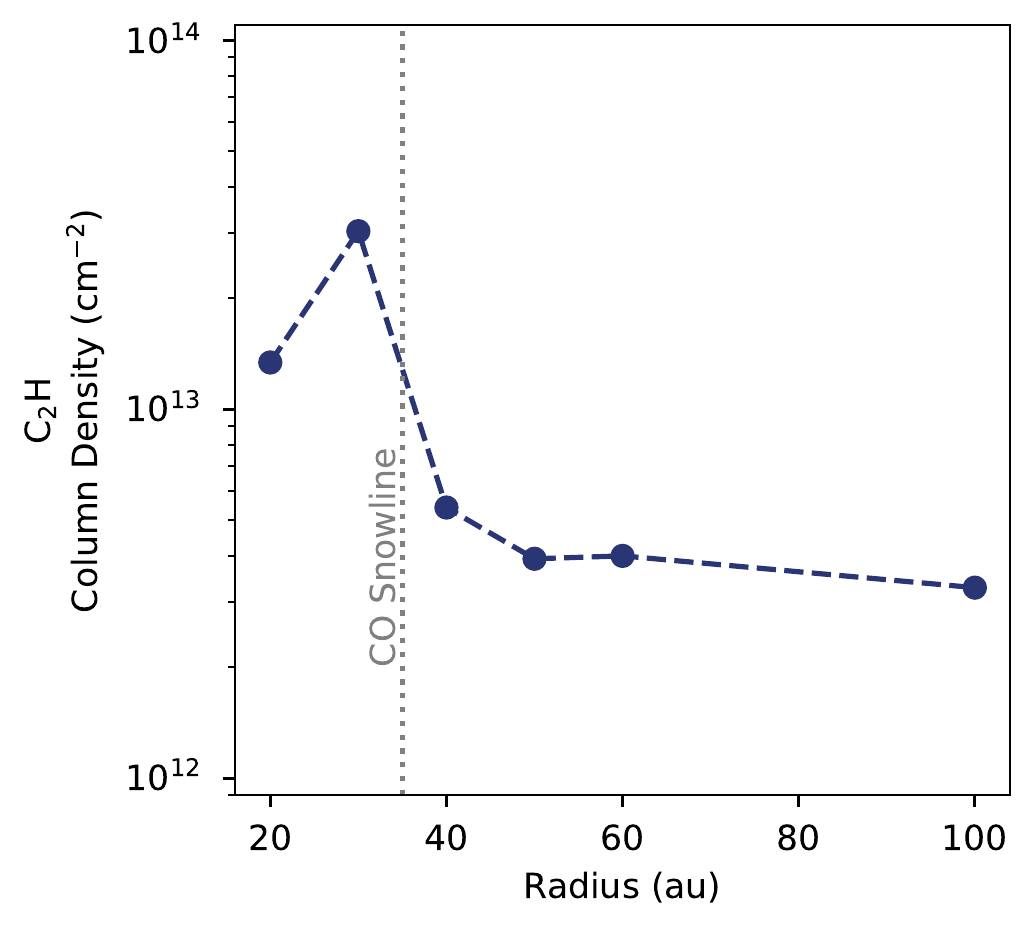}
    \caption{C$_2$H column density at different radial locations for the fiducial model. Modeled C$_2$H column densities remain 1-2 orders of magnitude below observed values of 10$^{14}$ - 10$^{15}$ cm$^{-2}$. }
    \label{fig:c2h_col_dens}
\end{figure}

In our models, the selective removal of oxygen through ice sequestration within the CO snowline enhances the C/O ratio to about unity.  This is still below the C/O ratios ($\gtrsim1.5$) inferred for disk surface layers via observations of C$_2$H \citep{2016ApJ...831..101B, 2021ApJ...910....3B}.   Indeed, in our fiducial model the C$_2$H column density peaks at 3$\times10^{-13}$ cm$^{-2}$ (Figure~\ref{fig:c2h_col_dens}) as compared to observed C$_2$H column densitiy peaks of $10^{14} - 10^{15}$ cm$^{-2}$ \citep{2019ApJ...876...25B, 2021arXiv210906391G}.  Moreover, we achieve a high C/O only interior to the CO snowline, whereas observations of C$_2$H emission probe disk regions beyond the snowline.  Thus, our model does not reproduce the carbon-rich chemistry observed in disk surface layers.

While C$_2$H is primarily a tracer of UV chemistry and C/O ratios \citep{2021ApJ...910....3B}, X-ray chemistry may further drive C$_2$H production in surface layers of the disk \citep{nomura2007, henning2010}. In our models we have included cosmic ray ionization, which behaves similarly to X-rays in mechanism (e.g. ionization of molecular hydrogen). X-ray induced chemistry can be important on the surface layers and can be considered in further implementations of the model.

In our model, CO destruction results in the formation of O-bearing and C-bearing species with comparable volatilities (e.g.H$_2$O and midsize hydrocarbons), which freeze out onto grains at similar temperatures.  CO remains the main gas-phase carrier of C and O, resulting in a gas-phase C/O of unity.  Unless our model over-predicts the formation of large carbon chains at the expense of more volatile carbon carriers like CH$_4$, a combined chemistry and pebble growth framework cannot explain the C/O levels $>$1 inferred from disk observations. 
Additional carbon insertion, i.e. from photoablation of grains \citep{2021ApJ...910....3B}, is then necessary to explain C/O ratios above unity. 

The carbon abundance in refractory grains is $10^{-4}\times 100\epsilon$, where $\epsilon$ is the dust to gas ratio \citep{2021ApJ...910....3B}. After 1 Myr of evolution at 30 au, dust is depleted by a factor of 50, such that $\epsilon = 2\times10^{-4}$, resulting in a carbon abundance in grains of $2\times10^{-6}$. The ablation of 100\% of these remaining small grains would be necessary to raise the C/O ratio to $\sim$2. This scenario is not realistic, implying that carbon grain destruction should be ongoing over the course of grain growth and disk evolution.  Additionally, grain ablation would provide similar feedbacks on the chemistry as grain growth, via the removal of UV opacity and the increase in gas-phase C/O.   Further implementations of this model will be able to capture grain ablation simultaneously with growth and chemistry.

\section{Conclusions}
We have developed a new technique to self-consistently model the growth and dynamics of small dust with a full gas and grain chemical network in a 1D vertical column of the protoplanetary disk. Using this new approach, we have explored in detail the influence of dust growth on CO depletion and C/O ratios in the disk. We find that a disk turbulence parameter of $\alpha = 10^{-3}$ is most closely able to match observed CO depletions of 10-100 below ISM abundances \citep{2013ApJ...776L..38F, 2019ApJ...883...98Z}. Additionally, we find that selective removal of H$_2$O ice inside of the CO snowline is able to increase C/O to unity, but are unable to reach higher C/O values implied by the observations of C$_2$H in disks \citep{2016ApJ...831..101B, 2021ApJ...910....3B}.

\section*{Acknowledgements}
This work was completed in part with resources provided by the University of Chicago’s Research Computing Center.

The authors acknowledge support from NASA’s Emerging Worlds Program, grant 80NSSC20K0333, and Exoplanets Research Program, grant 80NSSC20K0259.

J.B.B. acknowledges support from NASA through the NASA Hubble Fellowship grant \#HST-HF2-51429.001-A awarded by the Space Telescope Science Institute, which is operated by the Association of Universities for Research in Astronomy, Incorporated, under NASA contract NAS5-26555. 

\software{
\texttt{astrochem} \citep{2015ascl.soft07010M},
\texttt{matplotlib} \citep{2007CSE.....9...90H},
\texttt{mpi4py} \citep{2021CSE....23d..47D},
\texttt{numpy} \citep{2020Natur.585..357H}
}

\bibliography{bibliography.bib}

\appendix

\section{Description of Chemical Model}\label{sec:chem_model}
Our chemical model is built on top of the \texttt{astrochem} numerical solver, with added gas phase reactions including self-shielding and excited H$_2$. Gas-grain reaction rates are calculated following \citet{2011A&A...534A.132V} such that the desorption and hydrogenation rates for each species are limited by the species abundance in the ice and the number of monolayers allowed to desorb from the grain. Hydrogenation is limited to the reactions listed in table \ref{tab:hydrogenation} as in \citet{2011A&A...534A.132V}. These adjustments make our complete chemical solver more similar to the DALI code \citet{2012A&A...541A..91B} while maintaining the ease of use of the \texttt{astrochem} solver.

\begin{table}
    \centering
    \begin{tabular}{lllll}
    \hline
    \hline
        C $\ \rightarrow$ &CH $\ \rightarrow$ &CH$_2$ $\ \rightarrow$ &CH$_3$ $\ \rightarrow$ &CH$_4$ \\
        N $\ \rightarrow$ &NH $\ \rightarrow$ &NH$_2$ $\ \rightarrow$ &NH$_3$ \\
        O $\ \rightarrow$ &OH $\ \rightarrow$ &H$_2$O  \\
        S $\ \rightarrow$ &HS $\ \rightarrow$ &H$_2$S \\
    \hline
    \end{tabular}
    \caption{Hydrogenation reactions included in the network}
    \label{tab:hydrogenation}
\end{table}

\subsection{Gas phase reactions}

The majority of gas phase reactions follow directly from the UMIST database \citep{2013A&A...550A..36M}, from which the $\alpha$, $\beta$, and $\gamma$ coefficients for all reactions are taken. For reactions involving two gas species, the reaction rate constant, $k$ is given by the formula

\begin{equation}\label{eq:gas-phase}
    k = \alpha\left(\frac{T}{300\text{ K}}\right)^\beta \exp\left(\frac{-\gamma}{T}\right) \text{ cm}^3\text{s}^{-1}.
\end{equation}

Reactions involving cosmic rays have a rate coefficient given by

\begin{equation}
    k = \alpha\frac{\zeta}{\zeta_0}\frac{\gamma}{1-\omega}\left(\frac{T}{300\text{ K}}\right)^\beta\text{ s}^{-1}
\end{equation}
where $\zeta$ is the cosmic ray ionization rate and $\zeta_0 = 1.3\times10^{-17}$ is taken as the standard ionization rate. The FUV dust albedo, $\omega$ is assumed to be equal to 0.5 for all cosmic-ray reactions.

Modifications to the UMIST reaction rate library involve self-shielding and excited H$_2$ reactions. All reactions involving UV photons take the form

\begin{equation}
    k = \Theta \alpha \chi \exp\left(-\gamma A_v\right)\text{ s}^{-1}
\end{equation}
where $\chi$ is the unattenuated UV intensity and $A_v = \tau_{UV}/3.02$ is the visual extinction. The shielding factor, $\Theta$, is equal to one for all reactions except for photodissociation of CO and H$_2$. For these shielded reactions, the column densities of CO and H$_2$ above the given location is calculated and the self-shielding factor is determined from \citet{2009A&A...503..323V} for CO dissociation and \citet{1996ApJ...468..269D} for H$_2$.

Excited gas phase hydrogen reactions involve reactions with H$_2$ that has been vibrationally excited by UV radiation. This vibrational excitation allows the H$_2$ molecules to overcome reaction barriers of otherwise endothermic reactions. Although H$_2$ can be excited into a number of vibrational states, we follow \citet{2018A&A...615A..75V} and consider a separate species of H$_2$ excited to a weighted average vibrational pseudo-level of $\nu = 6$ or $E/k = 30163$~K, and spontaneous decay rate of $2\times10^{-7}$~s$^{-1}$. The rate of H$_2$ excitation is a combination of excitations from UV pumping, collisions with H$_2$ and H, and temperature. Similarly, the de-excitation rate is a combination of spontaneous decay, FUV pumping, and collisions. The rate of collisions, $q$, is given by the formula

\begin{equation}
    \log\left(\frac{q}{\text{cm}^3\text{s}^{-1}}\right) = a + \frac{b}{t} + \frac{c}{t^2}
\end{equation}
where $t = 1 + T/(1000\text{ K})$ and the coefficients $a$, $b$, and $c$, are fit from \citet{1999MNRAS.305..802L} for both H$_2$-H and H$_2$-H$_2$ collisions. The rate of excitation then is given by

\begin{equation}
    k_{ex} = 10k_{\text{diss}}+(q_{(H_2-H)}n_H+q_{(H_2-H_2)}n_{H_2})\exp\left(\frac{-30163\text{ K}}{T}\right)\text{ s}^{-1}
\end{equation}
where $k_{\text{diss}}$ is the rate of UV dissociation of H$_2$. Similarly the rate of de-excitation is given by

\begin{equation}
    k_{dex} = 10k_{\text{diss}}+(q_{(H_2-H)}n_H+q_{(H_2-H_2)}n_{H_2})+2\times10^{-7}\text{ s}^{-1}.
\end{equation}

Reactions between excited H$_2$ and other gas phase species follow a modified version of equation \ref{eq:gas-phase} to account for the vibrational energy of the excited H$_2$ molecule

\begin{equation}
    k = \alpha\left(\frac{T}{300\text{ K}}\right)^\beta \exp\left(\frac{-(\gamma-30163)}{T}\right) \text{ cm}^3\text{s}^{-1}.
\end{equation}

\subsection{Gas-grain reactions}

Gas-grain reactions for our modified network include freeze-out, thermal desorption, photodesorption, and hydrogentation. We follow the reaction rates of \citet{2011A&A...534A.132V}, however we allow for desorption of two monolayers at a time from the grain surface, doubling the number of binding sites available for desorption and hydrogenation reactions. Our reaction rates for thermal desorption, photodesorption, and hydrogentation follow equations (4), (6), and (7) of \citet{2011A&A...534A.132V} respectively. In our model, however, the total number of binding sites per grain is given by

\begin{equation}
    N_{b} = 4\pi a_{gr}^2N_{ss}N_{des}
\end{equation}
where $a_{gr}$ is the grain radius, $N_{ss} = 8\times 10^{14}\text{ cm}^{-2}$ is the number of binding sites per unit grain surface, and $N_{des} = 2$ is the number of layers allowed to desorb. Binding energies of ice species are taken from the UMIST12 database \citep{2013A&A...550A..36M}.

\section{Sensitivity to physical parameters}\label{appendix A}

\subsection{CO depletion\label{co param sensitivity}}

\begin{figure}
    \centering
    \includegraphics[width=0.7\linewidth]{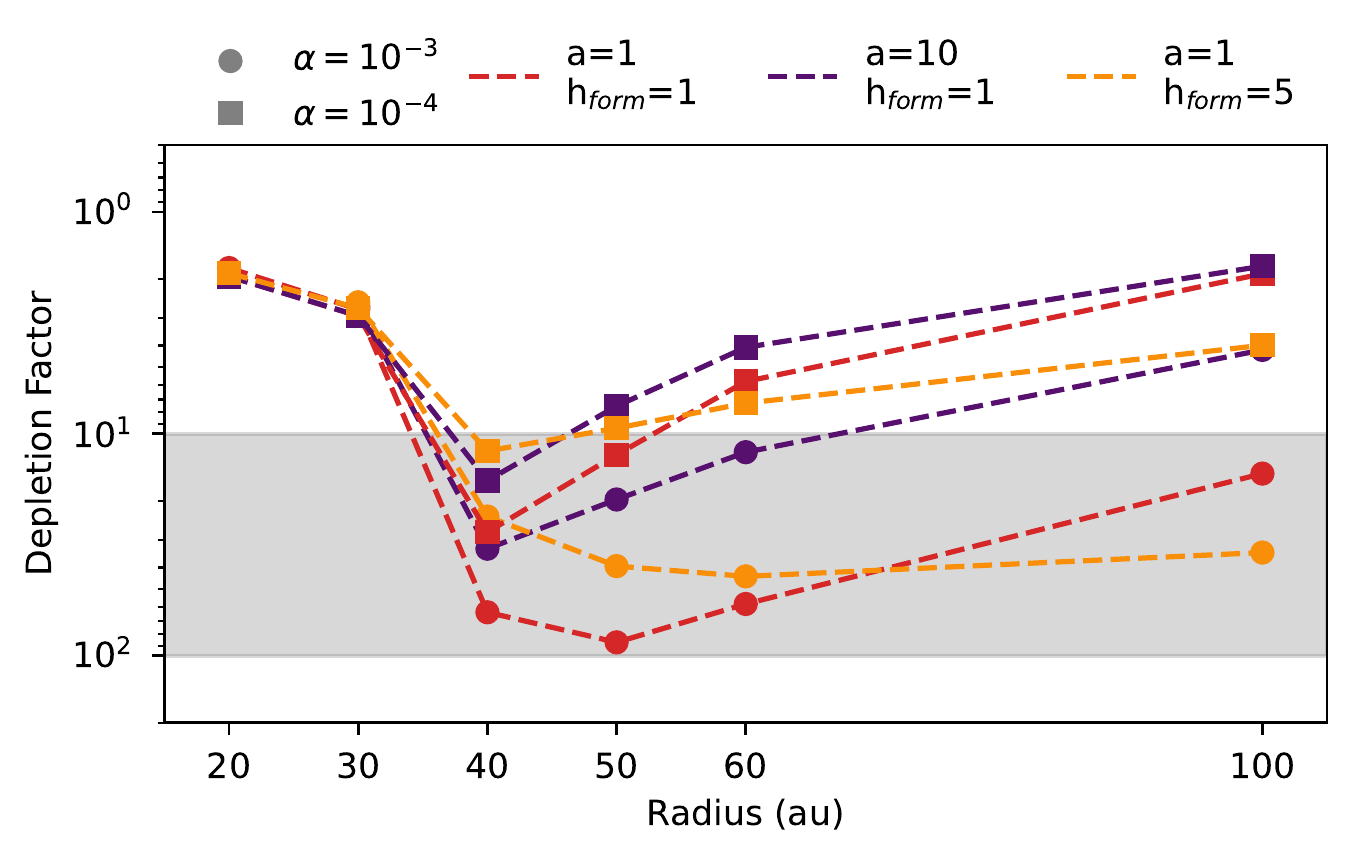}
    \caption{Depletion of carbon monoxide in the warm molecular layer (as in figure \ref{fig:co_depletion}) for various model values of turbulence ($\alpha$), pebble growth rate ($a$), and pebble formation region ($h_{form}$). Observed depletion factors of 10-100$\times$ below ISM values are indicated by the grey region as in figure~\ref{fig:co_depletion}.}
    \label{fig:co_depl_appendix}
\end{figure}

While depletion of CO outside of the snowline is a robust outcome of our model, the extent of the depletion depends on physical parameters of the disk, such as the turbulence, $\alpha$, pebble growth rate, $a$ (see Eq.~\ref{eq:growth_time}), and location of pebble formation $h_{form}$. The dependence of CO depletion on $\alpha$ is discussed in section 4, and we elaborate on the dependence of $a$ and $h_{form}$ here. Figure~\ref{fig:co_depl_appendix} shows the dependence of CO depletion on $a$ and $h_{form}$ for $\alpha$ values of $10^{-3}$ and $10^{-4}$.

The consequences of slower growth are shown by the dark purple dashed line in figure~\ref{fig:co_depl_appendix} as we increase the dust growth timescale, $a$, by an order of magnitude. As expected, slower dust growth leads to less loss of CO around the midplane.  Further, the slower growth leaves the disk more optically thick to penetrating radiation, resulting in a diminished level of photoprocessing in the upper layers of the disk.

To ensure that the details of depletion are not dependent on our choice of where we allowed pebbles to form, we also carried out a simulation where pebbles formed throughout the entire column of the disk (h$_{form}$=5, figure~\ref{fig:co_depl_appendix}, orange line).  In the cold outer disk regions CO remains frozen up to high elevations, so CO depletion is more efficient in this case.  By contrast, around the radial snowline location (40 au) the CO snow surface is closer to the midplane, so forming pebbles at higher elevations does not enhance CO depletion. In fact, the removal of dust grains bare of CO ice at the upper layers of the disk decreases the amount of CO that is eventually depleted via pebble growth.

\subsection{C/O ratios}\label{C/O param sensitivity}

\begin{figure}
    \centering
    \includegraphics[width=0.5\linewidth]{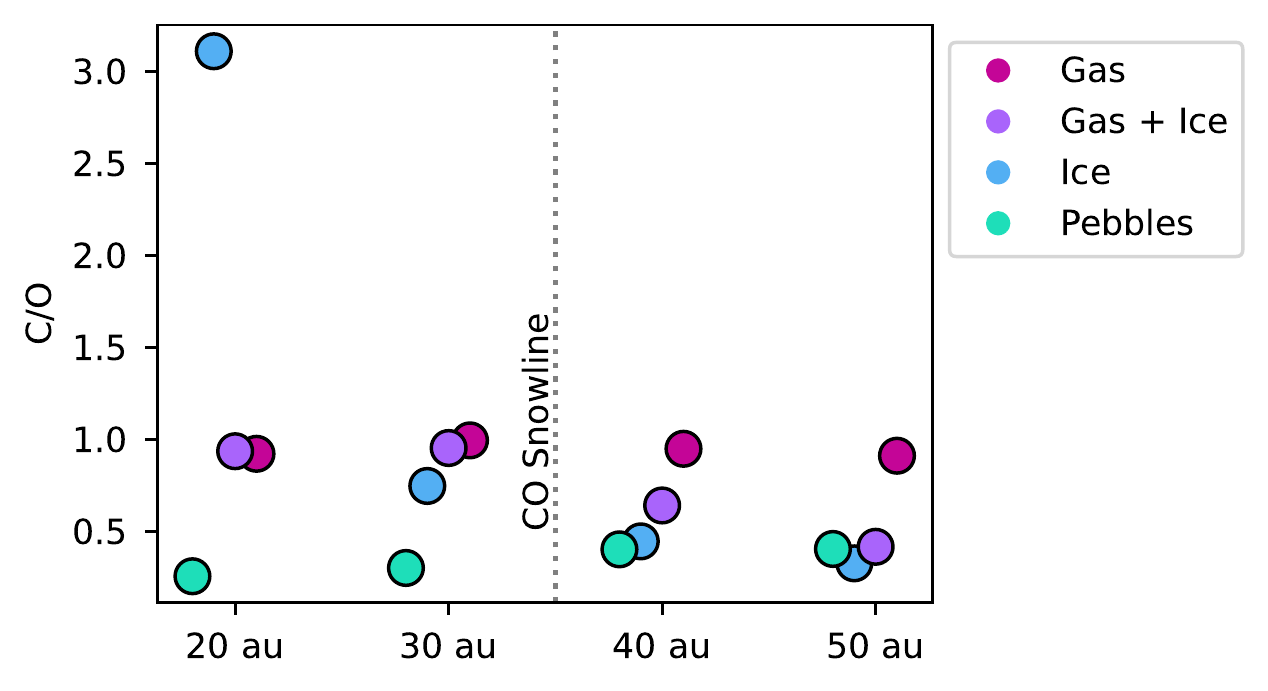}
    \caption{column averaged C/O ratios at 20, 30, 40, and 50 au. In general, the C/O ratios in the upper regions of the disk are the same as the bulk ``Gas + Ice" ratio, while the gaseous C/O is dominated by gas phase CO at the midplane.}
    \label{fig:CO_radial}
\end{figure}

Figure~\ref{fig:CO_radial} summarizes the column-averaged C/O ratios for different volatile components in the fiducial model.  Given the that the physical conditions in a disk will be critical in setting the chemical evolution that occurs, we next explore whether changes to disk conditions may allow for higher C/O ratios to be reached than found here.  We focus primarily on the conditions around 30 au as this is where the highest C/O ratios were found in our model, but note most observations at millimeter wavelengths are not sensitive to regions within the CO snowline.  Nonetheless, our goal is to find what conditions allow high C/O ratios to develop in the most favorable region of the disk.
We present the column averaged gas, ice, and total (gas + ice) C/O ratios and note that in the upper layers of the disk dust grains are largely devoid of icy mantles, and the conditions there are set by the gas.

\subsubsection{Cosmic-ray ionization}

\begin{figure}
    \centering
    \includegraphics[width=0.5\linewidth]{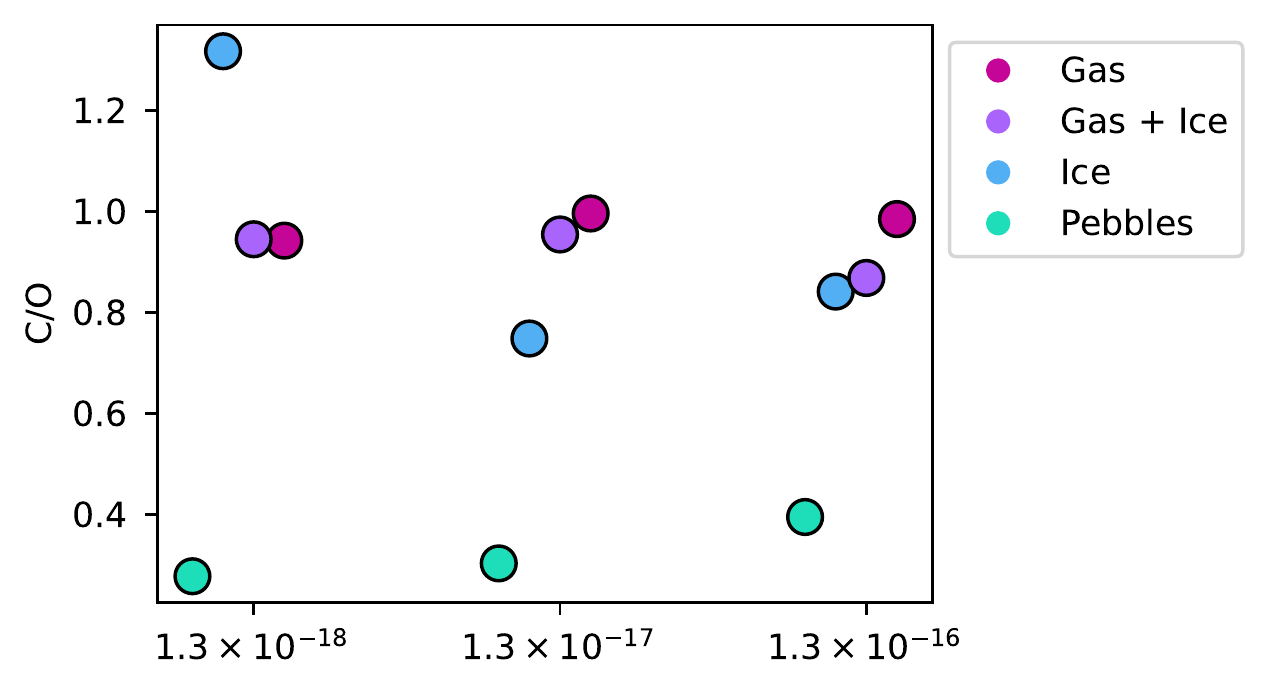}
    \caption{Column averaged C/O ratios after 1 Myr at 30 au for differing cosmic ray ionization rates in $s^{-1}$. Although we achieve C/O$>1$ for the lowest ionization rate, the abundance of ice species is very low, and the combined gas and ice C/O remains below unity.}
    \label{fig:cr_CO_scatter}
\end{figure}

Cosmic rays are important drivers of chemistry in the outer regions of protoplanetary disks.  The effects of different cosmic ray ionization rates on C/O ratios are shown in figure~\ref{fig:cr_CO_scatter}, which again plots the column averaged C/O ratios for the gas, ice, total (gas+ice), and pebbles at 30 au, assuming different cosmic ray ionization rates. As the ionization rate increases, the C/O ratio of the pebbles increases slightly.  
This is because the secondary carbon products are available at earlier times, allowing them to be incorporated into forming pebbles. The lower ionization rates lead to slower chemistry, and the products of cosmic-ray driven reactions are formed after pebble growth has peaked.

At higher ionization rates, the removal of secondary carbon products via pebble growth leaves the remaining total column with a smaller C/O ratio. The total C/O value is very close to the gas value for the lowest ionization rate, and to the ice value for the highest ionization rate, as  high ionization rates drive reactions that preferentially remove elements from gaseous molecules and incorporating them into less volatile ices.
However, the total C/O ratio varies by only $\sim$10\% over  a range spanning three orders of magnitude in the ionization rate, suggesting that the resulting ratio is not highly sensitive to this parameter.

\subsubsection{UV radiation}

\begin{figure}
    \centering
    \includegraphics[width=0.5\linewidth]{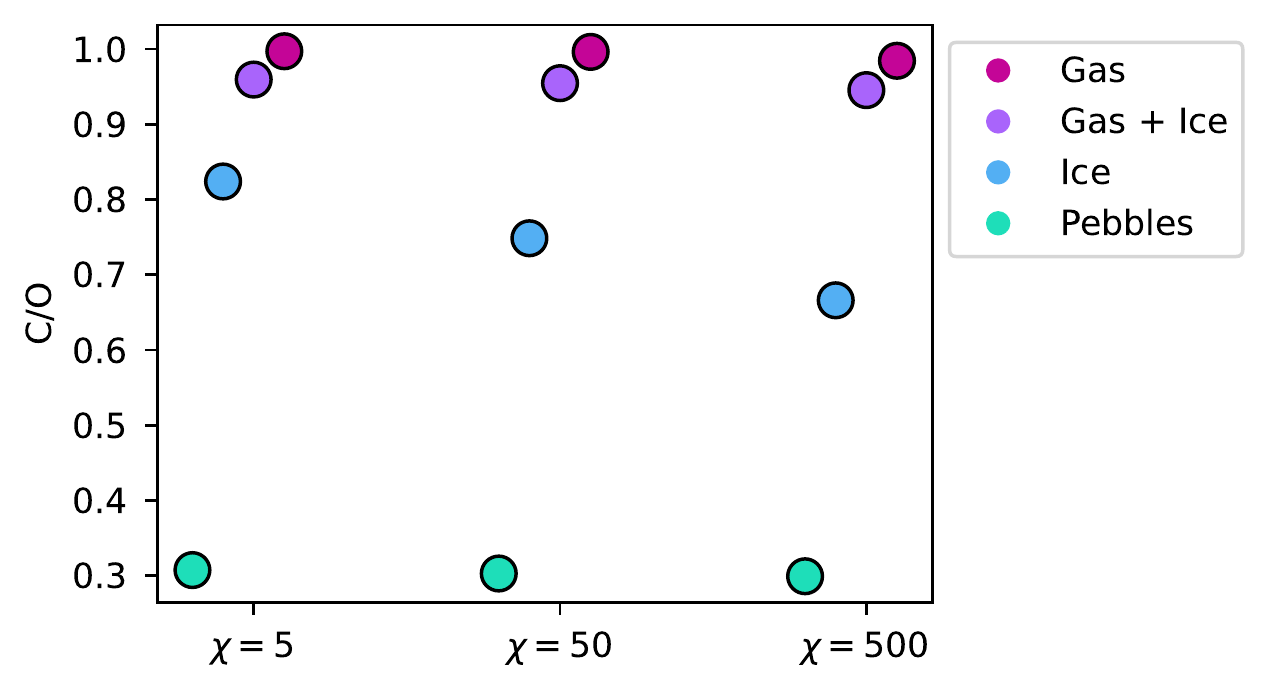}
    \caption{C/O ratios after 1 Myr evolution with UV fields at 5, 50, and 500 times interstellar value in Draine units}
    \label{fig:uv_CO_scatter}
\end{figure}

The amount of photoprocessing in the PDL is set, in part, by the intensity of the UV radiation field.
We varied the ambient field from $\chi = $ 5 and 500 Draine units, representing a factor of 10 decrease and increase from the fiducial $\chi = 50$. The results are shown in Fig~\ref{fig:uv_CO_scatter}, which again shows the C/O ratio among the different volatile components in the column at the end of the simulations.  The differences are relatively small, particularly for the pebbles and gas.  Thus, the incident UV field does not strongly impact the composition of material incorporated into planetesimals at the midplane. Indeed, the midplane is largely protected from UV radiation, and most of the pebble mass forms before photoprocessed material is mixed down to the midplane.  

\subsubsection{Growth \& dynamics}
\begin{figure}
    \centering
    \includegraphics[width=0.5\linewidth]{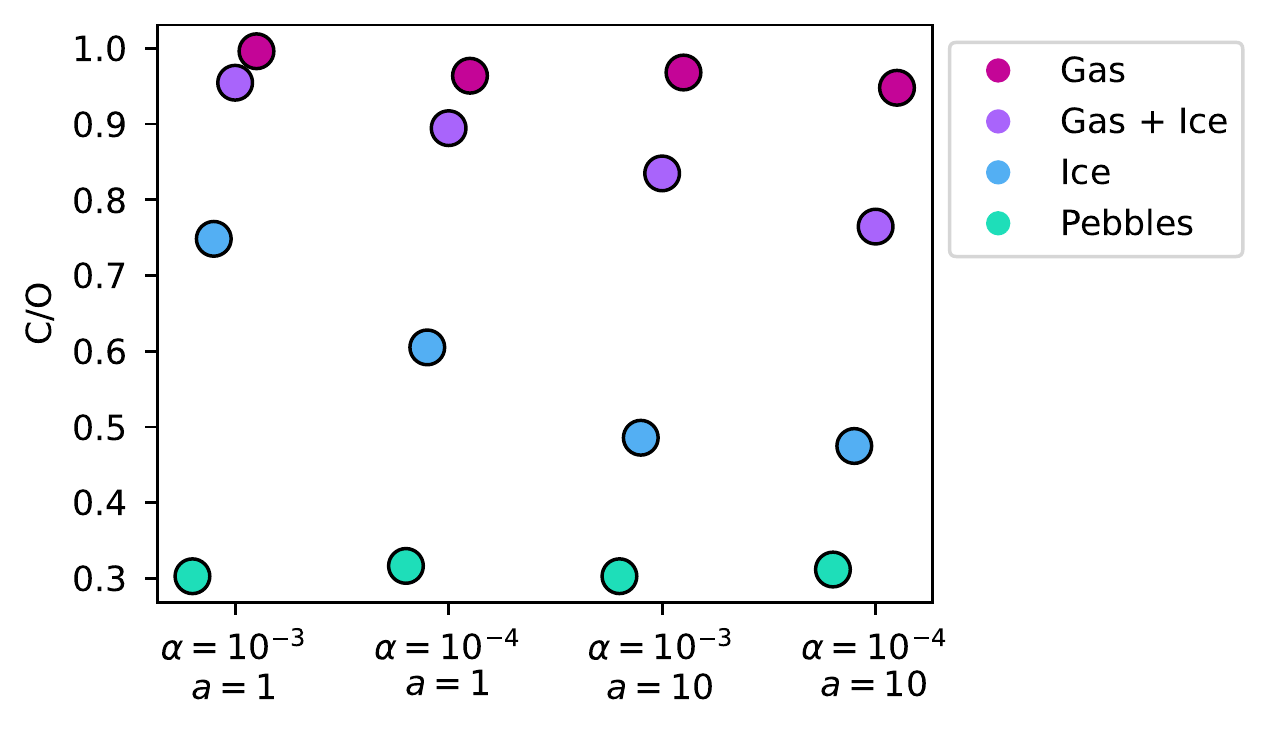}
    \caption{C/O values for different alpha and growth timescale parameters of the disk. Decreasing alpha and increasing the growth timescale both have the net effect of lower the C/O ratio at 1 Myr of evolution.}
    \label{fig:CO_phys_params}
\end{figure}

As discussed above, disk chemical evolution is also controlled by the amount of dust present and the level of mixing, as dust grain abundances set the level to which photons can penetrate inside of the disk and the available sites for gas-grain chemistry to occur.  As such, we altered the dust growth time scale and level of turbulence to evaluate how they impacted the C/O ratios.   Figure~\ref{fig:CO_phys_params} shows the C/O ratios for $\alpha$ values of $10^{-3}$ and $10^{-4}$, and growth timescale parameters of $a=1$ (rapid) and $a=10$ (slower).

Increasing the growth timescale ($a$ in equation~\ref{eq:growth_time}) slows the removal of small dust from the column. This has the effect of 1) slowing the removal of oxygen via ice sequestration, and 2) slowing dust depletion and thus inhibiting photoprocessing in the PDL.
These two effects together keep the gas C/O ratio near unity as CO remains abundant in the column.  Meanwhile, the slow removal of oxygen rich ices means that the C/O ratio in the ice is lower, and that the ice remains a more significant  component of the total volatile budget in the column.  This is reflected in the lower total (gas + ice) C/O ratio of the $a$=10 case.

Decreasing the the level of turbulence from $10^{-3}$ to $10^{-4}$ slows the diffusion of material from the upper regions of the disk to the pebble forming midplane. Because small grains are less efficiently mixed downwards, optical depths remain high in upper regions of the disk. Additionally, H$_2$O on these grains is not carried downward as quickly to the pebble forming region, resulting in less removal of H$_2$O and, by extension, oxygen. These effects combine to reduce the total C/O ratio in the disk at 1 Myr compared to the $\alpha$=10$^{-3}$ case, with gas, ice, and bulk C/O ratios below unity.

\end{document}